\documentclass[]{JHEP3}
\usepackage{epsfig}
\input axodraw.sty
\newcommand{\mett}{{\not\!\!E}_{T}}
\newcommand{\rpv}{{\not\!\!R}_{P}}

\def\beq{\begin{equation}}
\def\eeq{\end{equation}}
\def\bea{\begin{eqnarray}}
\def\eea{\end{eqnarray}}

\newcommand{\threegraphs}[3]{%
\unitlength=1in
\begin{picture}(5.4,5)
\put(1.35,0){\epsfig{file=#3.eps, width=2.7in}}
\put(0,2.5){\epsfig{file=#1.eps, width=2.7in}}
\put(2.7,2.5){\epsfig{file=#2.eps, width=2.7in}}
\put(1.35,2.3){(c)}
\put(0,4.8){(a)}
\put(2.7,4.8){(b)}
\end{picture}}

\newcommand{\twographs}[2]{%
\unitlength=1in
\begin{picture}(5.4,2.7)
\put(2.7,0){\epsfig{file=#2.eps, width=2.7in}}
\put(0,0){\epsfig{file=#1.eps, width=2.7in}}
\put(0,2.7){(a)}
\put(2.7,2.7){(b)}
\end{picture}}

\newcommand{\vtwographs}[2]{%
\unitlength=1in
\begin{picture}(4,6)
\put(0,0){\epsfig{file=#2.eps, width=4in}}
\put(0,3){\epsfig{file=#1.eps, width=4in}}
\put(-0.2,6){(a)}
\put(-0.2,3){(b)}
\end{picture}}

\newcommand{\fourgraphs}[4]{%
\unitlength=1in
\begin{picture}(5.4,5)
\put(0,0){\epsfig{file=#3.eps, width=2.7in}}
\put(2.7,0){\epsfig{file=#4.eps, width=2.7in}}
\put(0,2.5){\epsfig{file=#1.eps, width=2.7in}}
\put(2.7,2.5){\epsfig{file=#2.eps, width=2.7in}}
\put(0,2.3){(c)}
\put(0,4.8){(a)}
\put(2.7,4.8){(b)}
\put(2.7,2.3){(d)}
\end{picture}}

\preprint{CERN-TH/2001-360}

\title{Investigating the Supersymmetric Explanation of Anomalous CDF lepton(s)
photon(s) Missing-\boldmath{$E_T$} Events}

\author{B.C.  Allanach, S. Lola and K. Sridhar\footnote{On leave of absence 
from the Tata Institute of Fundamental Research, Homi Bhabha Road, 
Mumbai 400 005, India.} \\
 \noindent CERN, Geneva 23, CH-1211, Switzerland }


\abstract{The recent excess over the Standard Model prediction 
in the 
 $\mu \gamma$
missing-$E_T$ ($\mett$) channel 
reported by CDF can be well-explained by resonant smuon production
with a single dominant R-parity violating coupling $\lambda'_{211}$,
in the context of models where
the gravitino is the lightest supersymmetric particle.
The slepton decays to the lightest neutralino
and a muon followed by neutralino decaying to a gravitino and photon.
The kinematical distributions are fitted well by our hypothesis and we use
them to constrain the available parameter space.
The model also provides an explanation for 
the $ee\gamma\gamma\mett$ event observed in Run I of the
Tevatron by the CDF experiment. 
Our model predicts an excess of between 5 and 35 events in a $\gamma \mett$
channel at Run I.
We provide predictions for signatures
expected by the model at run II. 
}

\begin{document}

\section{Introduction}
In spite of the remarkable agreement of the Standard Model (SM) with available
data from high-energy experiments, it is expected to be only a low-energy 
manifestation of a more complete theory at energy scales beyond a TeV. This 
new TeV-scale physics is expected to ameliorate the problems that beset the 
SM because of the huge discrepancy between the electroweak scale and the 
Planck (or GUT) scale. The most popular candidate for such an extension of the
SM has been its supersymmetric generalisation which, in its
simplest form, is the Minimal Supersymmetric Standard Model (MSSM). 
The gauge structure of the MSSM essentially replicates that of the SM but,
in the Yukawa sector, in addition to the usual Yukawa couplings of the fermions
to the Higgs (responsible for the fermion masses), other interactions
involving squarks or sleptons are possible.
 
The relevant part of the superpotential containing the Yukawa interactions
involving squarks or sleptons is given in terms of the chiral superfields by
\beq
W_{RPV}=\frac{1}{2}\lambda_{ijk} L_iL_j{\bar E}_k+\lambda'_{ijk}L_iQ_j{\bar D_k}+
\frac{1}{2}\lambda''_{ijk}{\bar U_i}{\bar D_j}{\bar D_k}+\mu_i L_i H_2
\label{eq:superpot}
\eeq
where $L$ $(Q)$ are the left-handed lepton (quark) superfields while ${\bar
E},{\bar D},$ and ${\bar U}$ contain the corresponding right-handed fields,
and $i,j,k$ generation indices. $\lambda$ and $\lambda'$ are lepton-number 
($L-$) violating, the $\lambda''$ couplings are baryon-number ($B-$) 
violating and the last term is a $L$-violating bilinear coupling.
The simultaneous existence of the $L-$ and $B-$-violating couplings
can induce a catastrophically high rate for proton decay and are usually
forbidden in the MSSM by invoking a discrete symmetry called $R$-parity 
where $R=(-1)^{(3B+L+2S)}$, where $S$ is the spin of the particle, so that 
the SM particles have $R=1$, while their superpartners have $R=-1$.
However, $R$-conservation is too strong a requirement to avoid the unwanted
proton decay for it can be effectively forbidden assuming that 
either the $L$-violating or the $B$-violating couplings 
in Eq.~\ref{eq:superpot} 
are present, but not both. Limits on the $R$-violating couplings derived 
from existing experimental information have been summarised in 
Ref.~\cite{rparrev}.

In the presence of $R$-violating couplings, the lightest supersymmetric 
particle (LSP), which is usually the neutralino, is $not$ stable and can decay
through $R$-violating modes \cite{pheno}.
This is in contrast to the $R$-conserving MSSM
where the LSP is stable and this stability is a very desirable feature if 
the LSP were to be a viable dark matter candidate. In the $R$-violating
case, the neutralino cannot be a dark matter candidate unless the $R$-violating
couplings are very small so as to ensure that the lifetime of the neutralino is
much more than the age of the universe. The situation can be saved, however,
in theories where the gravitino (the spin-3/2 superpartner of the graviton) 
is the lightest supersymmetric particle: a circumstance that can be realised 
very naturally in theories with gauge-mediated supersymmetry breaking 
\cite{gmsb}. The light gravitino is long-lived enough to account for dark 
matter (or, at least, the hot component of dark matter) even in the presence of $R$-violating couplings \cite{bmy}.

\section{Physics of Light gravitinos}
Even though gravity is naturally incorporated if supersymmetry is realised as a 
gauge symmetry, the gravitational sector is usually irrelevant for collider
phenomenology because of the feebleness of gravitational interactions. But if
supersymmetry is broken spontaneously, the gravitino acquires a mass by
absorbing the would-be goldstino and in the high-energy limit the gravitino
has the same interactions as the goldstino \cite{light-gra}. These 
interactions are proportional
to $1/m_{\tilde G}$ and consequently the interactions of the gravitino 
can become important for processes at collider energies in the $m_{\tilde G}
\rightarrow 0$ limit. The mass of the gravitino is related to $F_0$, 
the fundamental scale-squared of supersymmetry breaking, by the following
relation: 
\beq
m_{\tilde G} = {F_0 \over \sqrt 3 M_P} .
\eeq
$M_P= 2.4 \times 10^{18}$~GeV is the reduced Planck mass, and using this
value one obtains 
\beq
m_{\tilde G} = 5.9 \times 10^{-5} {F_0 \over ({500 {\rm GeV}})^2} {\rm eV} .
\eeq
Given a lower bound on the value of $F_0$ one can then deduce a lower bound 
on the mass of the gravitino which, in turn, yields a bound on the interactions
of the gravitino with the SM particles.

To make these considerations more concrete, we write down the relevant part of
the supersymmetric Lagrangian containing the gravitino interactions:
\beq
{\cal L} = {1 \over 8M_P} \bar \lambda^A\gamma^{\rho}\sigma^{\mu\nu} 
\tilde G_{\rho}F_{\mu\nu}^A+{1 \over \sqrt{2} M_P} \bar \psi_L\gamma^\mu
\gamma^\nu\tilde G_{\mu}D_{\nu}\phi + {\rm h.c.} ,
\label{e1}
\eeq
where $\tilde G$ is the gravitino field, $\lambda^A$ the gaugino field, 
$F_{\mu\nu}^A$  the corresponding field strength and $(\phi, \psi)$ the
scalar and the fermionic components of the chiral supermultiplets. At
the level of a effective interaction, the spin-3/2 gravitino field can be 
well described by its spin-1/2 goldstino component when it appears as
an external state, i.e.
\beq
\tilde G_\mu=\sqrt{2 \over 3} {i \over m_{\tilde G} } \partial_\mu\tilde G .
\eeq
Using this limit in Eq.\ref{e1}, allows one to compute the decay widths
of the process $\chi_i \rightarrow \gamma/Z \tilde G$, for example. These are:
\begin{eqnarray}
\Gamma (\chi^0_i \rightarrow \gamma \tilde G) &=& {\kappa_{i\gamma} \over 48 
\pi} {m_{\chi^0_i}^5 \over M_P^2 m_{\tilde G}^2} \nonumber \\
\Gamma (\chi^0_i \rightarrow Z \tilde G) &=& {2 \kappa_{iZ_T} + \kappa_{iZ_L}
 \over 96 \pi} {m_{\chi^0_i}^5 \over M_P^2 m_{\tilde G}^2} \biggl \lbrack 1 - 
{m_Z^2 \over m_{\chi^0_i}^2} \biggr \rbrack ^4 ,
\end{eqnarray}
where 
\begin{eqnarray}
\kappa_{i \gamma} &=& \vert N_{i1} {\rm cos} \theta_W + N_{i2} {\rm sin} 
\theta_W \vert^2 \nonumber \\
\kappa_{i Z_T} &=& \vert N_{i1} {\rm sin} \theta_W - N_{i2} {\rm cos} 
\theta_W \vert^2 \nonumber \\
\kappa_{i Z_L} &=& \vert N_{i3} {\rm cos} \beta - N_{i4} {\rm sin} 
\beta \vert^2 .
\end{eqnarray}
The $N_{ij}$ are the $\chi^0_i$ components in standard notation.
The neutralino can also decay into a gravitino and a neutral Higgs particle
and the corresponding expressions for these are 
\beq
\Gamma (\chi^0_i \rightarrow \phi \tilde G) = {\kappa_{i\phi} \over 96 
\pi} {m_{\chi^0_i}^5 \over M_P^2 m_{\tilde G}^2} \biggl \lbrack 1 - 
{m_{\phi}^2 \over m_{\chi^0_i}^2} \biggr \rbrack ^4 ,
\eeq
where the Higgsino components are given by
\begin{eqnarray}
\kappa_{i h^0} &=& \vert N_{i3} {\rm sin} \alpha - N_{i4} {\rm cos} 
\alpha \vert^2 \nonumber \\
\kappa_{i H^0} &=& \vert N_{i3} {\rm cos} \alpha + N_{i4} {\rm sin} 
\alpha \vert^2 \nonumber \\
\kappa_{i A^0} &=& \vert N_{i3} {\rm sin} \beta + N_{i4} {\rm cos} 
\beta \vert^2 .
\end{eqnarray}
From the above equations, it is easy to convince oneself that the decay 
modes into the photon dominates over the decays into the $Z$ or the neutral 
Higgs boson because the decays into the latter states are phase-space
suppressed. If the neutralino is bino-dominated, then the branching ratio
into a photon and gravitino is nearly 100\%. 

\section{The CDF anomaly}
CDF has recently presented results 
on the production of combinations
involving at least one photon and one lepton ($e$ or $\mu$)
in $p{\bar p}$ collisions at $\sqrt{s}= 1.8 ~{\rm TeV}$,
using 86.34 pb$^{-1}$ of Tevatron 1994-95  data \cite{CDF}.
In general the  results were consistent with the standard model,
however 16 photon-lepton events with
large ${{\not\!\!E}_{T}}$ were observed, with
$7.6\pm0.7$ are expected.
Moreover, 11 of these events involved muons
(with 4.2 $\pm$ 0.5 expected)
and only 5 electrons (with 3.4 $\pm$ 0.3 expected), therefore there is a clear
asymmetry, which indicates the existence of a lepton flavour-violating process
involving muons.

As we proposed in an earlier paper \cite{us}, we suggest that the excess
can be simply understood in terms of smuon resonance production via
an $L$-violating $\lambda'$ coupling\footnote{Smuon resonances at hadron
colliders have been previously  
studied in a different context \cite{previous}.}
which decays predominantly into
a bino-dominated neutralino and a muon, with the neutralino further decaying
into a photon and a gravitino. The production and decay has
been shown in the Feynman diagram in Figure~\ref{feyn}. The merit of this
model that we proposed is that it is a natural explanation of the 
characteristics of the CDF anomaly: 1) the flavour-dependence is a direct
consequence of the $R$-violating coupling and, 2) the fact that the excess is
seen in final states involving photons emerges very neatly in the model 
because the decay $\chi_1^0 \rightarrow \gamma \tilde G$ dominates 
overwhelmingly over other decay modes. We note that both $R$-parity
violation and the existence of a very light gravitino are needed to explain the anomaly, in our model. Nonetheless, we emphasise that if one has
$R$-parity violating 
supersymmetry, a light gravitino is preferable
from dark matter considerations, as explained in the previous 
section.

A light gravitino has also been previously invoked \cite{kane} to explain
the $ee \gamma \gamma \mett$ CDF event~\cite{abe}, detected in searches
for anomalous production of missing transverse energy ($ \mett > 12$ GeV),
in events containing two isolated, central photons. The event was explained
in terms of the $R$-conserving production of a pair of selectrons and the
subsequent decay of each of these selectron into a $\gamma \tilde G$ final 
state. It has been shown \cite{baer} that this explanation is excluded in 
the framework of the minimal uni-messenger gauge mediated supersymmetry
breaking (GMSB) model, because of the
anomalously large rates for jets $+ \gamma + \mett$ events predicted by
this model. The problem can be traced back to the mass spectrum of the
uni-messenger models: in this version of GMSB models the charginos and 
second-lightest neutralinos are light which lead to large jets 
$+ \gamma + \mett$ rates not seen in experiment. However, in multi-messenger 
models of GMSB the charginos and the second-lightest neutralinos are heavier 
and one can have a viable explanation of the CDF event in these models
which is not in conflict with other existing experimental information.
For our purposes,
again a light neutralino of 100 GeV mass and a reasonably light smuon
in the mass range of about 150 GeV is needed but we require all the
other supersymmetric particles to be very massive. In the present
paper, we perform detailed fits to the experimentally measured distributions
of the anomalous events in order to determine the masses of the lightest
neutralino and the smuon with the assumption that all the other masses
are heavy enough not to be produced at the Tevatron. 
We do not attempt to place our scenario in the context of some specific 
model of GMSB, but point out that this is indeed possible in the case of 
multi-messenger models of GMSB. A detailed model-dependent study is relegated
to a later publication.

\FIGURE{
\label{feyn}
\begin{picture}(400,170)
\ArrowLine(60,10)(120,50)
\ArrowLine(120,50)(60,90)
\DashLine(120,50)(180,50){5}
\ArrowLine(180,50)(240,10)
\DashLine(180,50)(240,90){5}
\DashLine(240,90)(300,70){5}
\Photon(240,90)(300,130){5}{4}
\put(65,5){$q$}
\put(65,90){$\bar q'$}
\put(245,10){$\mu$}
\put(210,80){$\chi_0$}
\put(150,65){$\tilde\mu$}
\put(305,65){$\tilde G$}
\put(305,125){$\gamma$}
\end{picture}
\caption{Feynman diagram of resonant smuon production followed by neutralino decay.}
}

\section{Constraints}
If the anomalous events seen by the CDF experiment are to be attributed to
the production of a smuon resonance involving an $R$-violating operator,
we  can ask what the precise form of this operator is. To get a 
substantial cross-section for the production of the smuon resonance
one needs to couple it to valence quarks in the initial state.
This observation is then sufficient to specify the $R$-violating operator to 
be $L_2Q_1\bar{D}_1$ corresponding to the coupling $\lambda'_{211}$. This 
operator generates the interactions
${\tilde \mu}u\bar{d}$ and ${\tilde \nu}_\mu d \bar{d}$ (and charge 
conjugates), along with other supersymmetrised copies involving squarks. 
Therefore, if we invoke this operator to explain the CDF anomaly we will
simultaneously predict effects in other channels which will manifest itself
through the production of either sneutrinos or squarks. In our model, since 
we take the squarks to be heavy, their effects on experimental observables
will be negligible. On the other hand, the sneutrinos are necessarily
relatively light and can be
produced resonantly and should lead to observable effects in experimental 
situations. In the present paper, we not only analyse the smuon resonance
production in the context of the CDF anomaly but also provide predictions
for both the smuon and the sneutrino channels at Run I and Run II of the
Tevatron. The smuons, sneutrinos and the lightest neutralino are all light 
enough to be pair-produced through $R$-conserving channels. We also provide 
predictions for these cross-sections at Run I and Run II. 

For our analysis, we have essentially four parameters at our disposal: the
gravitino mass, $m_{\tilde G}$, the bino mass parameter $M_1$, the smuon
mass parmeter, $m_{\tilde \mu}$ and the $R$-violating coupling $\lambda'_{211}$.
The coupling, $\lambda'_{211}$, is constrained from  
$R_\pi = \Gamma (\pi \rightarrow e \nu) / (\pi \rightarrow \mu \nu)$
\cite{bgh} to be 
$< 0.059 \times \frac{m_{\tilde{d_R}}}{100 ~{\rm GeV}}$ \cite{rparrev}.
We note that the constraint involves a squark mass which is large in our
model. So the constraint from $R_\pi$ for our purposes is not very
relevant. However, instead of simultaneously fitting the four parameters
using the experimental data, we choose to work with fixed values of the
$\lambda'_{211}$ and $m_{\tilde G}$ and perform fits in $M_1$ 
and $m_{\tilde \mu}$. While the production of the smuon resonance is
through the $R$-violating mode, its decay needs to go through the 
$R$-conserving channel to a neutralino and muon final state. The $R$-violating
decay of the slepton is possible but constrained, in principle, by the
Tevatron di-jet data \cite{cdfjets} which exclude a $\sigma . B> 1.3 \times 
10^4$ pb at 95\%~C.L. for a resonance mass of 200 GeV. However, in practice
this does not provide a restrictive bound upon our scenario as long as 
the $R$-violating coupling is sufficiently small. We also add that the di-jet
bound is not very restrictive because it suffers from a huge QCD background.
By restricting $\lambda'_{211}$ to be small, we also avoid the possible 
$R$-violating decays of the $\chi_1^0 \rightarrow \mu jj$ or 
$\chi_1^0 \rightarrow \nu jj$ final states. With these considerations in 
mind we choose $\lambda'_{211}=0.01$. The gravitino mass is also fixed 
at $10^{-3}$ eV in our fits. We will discuss the effects of varying
the gravitino mass and the $R$-violating coupling later in this paper.

\section{Defining the model}

The supersymmetric model parameters that are relevant for
our discussion, are:
$M_1$, $M_2$, $\mu$, $\tan\beta$ and $m_0$, which determine
the chargino, neutralino and sfermion masses at low energies.
Since no other exotic cascade decays at CDF, 
are observed, we assume that:

$\bullet$
Charginos and other superparticles (except the slepton and the 
lightest neutralino)
are too heavy to be produced at the current energies.

$\bullet$
The decays of the lightest neutralino to gauge bosons
other than the photon are coupling and/or phase-space 
suppressed. 

These considerations constrain the
allowed supersymmetric parameter space, 
which as we are going to show, still has some generality within it.
To see this, let us look at the formulae that give the chargino
and neutralino masses and mixings in terms of the fundamental supersymmetric
parameters.

The tree-level neutralino mass matrix in the 
$\psi^0_j (-i \tilde{B},-i \tilde{W_{3}},\tilde{H}^{0}_{1},\tilde{H}^{0}_{2}) $
basis are the mass eigenstates of the matrix
\begin{equation}
 Y=\left(
    \begin{array}{cccc}
     M_{1}  &  0     & -m_{Z}\sin\theta_{W}\cos\beta 
                                    &  m_{Z}\sin\theta_{W}\sin\beta \\
     0      &  M_{2} &  m_{Z}\sin\theta_{W}\cos\beta
                                    & -m_{Z}\cos\theta_{W}\sin\beta \\
    -m_{Z}\sin\theta_{W}\cos\beta      
            &  m_{Z}\cos\theta_{W}\cos\beta
                    &   0           & -\mu                          \\
     m_{Z}\sin\theta_{W}\sin\beta   
            & -m_{Z}\cos\theta_{W}\sin\beta
                    & -\mu          &  0                            \\
\end{array}
\right)
\end{equation}
and are defined by
$ \chi^0_i = N^{ij} \psi^0_j$, with $N_{ij}$ being the unitary 
matrix which diagonalises $Y$. The respective mixings
in the basis $(\tilde{\gamma},\tilde{Z})$
instead of $(\tilde{B}, \tilde{W}_3)$, are given by
$N'_{j1} = N_{j1} \cos \theta_W + N_{j2} \sin \theta_W$,
$N'_{j2} = -N_{j1} \sin \theta_W + N_{j2} \cos \theta_W$,
$N'_{j3} = N_{j3}$ and $N'_{j4} = N_{j4}$. 
Finally, the respective chargino mass matrix in the 
$(\tilde{W}^{\pm}, \tilde{H}^{\pm})$ basis
is
\begin{equation}
 X = \left(   \begin{array}{cc}
                   M_{2}                   & m_{W}\sqrt{2}\sin\beta  \\
                 m_{W}\sqrt{2}\cos\beta  & \mu
                \end{array}    \right)
\end{equation}

In our work we will not make use of the
GUT inspired relation $ M_1 = \frac{5}{3} \tan^2 \theta_W M_2 $,
but will instead keep $M_1$ and $M_2$ generic.
As we see from the above formulas, a light neutralino can arise
either via a light $M_1$, a light $M_2$, or a light $\mu$. In the second 
and third cases
however, the chargino is also going to be light enough to be seen 
in cascade decays, which is not the case in CDF data.
Moreover, constraining the relative masses of $M_1$ and $M_2$ 
roughly determines the
photino component of the lightest neutralino.
Under these conditions $\tan\beta$ is expected to play a
relatively moderate role.
Table~\ref{tttt} contains the lightest chargino and neutralino masses,
and the magnitude of the photino-component of the
lightest neutralino, for different model parameters.
In this section only, we constrain ourselves to small and intermediate values
of 
$M_2$ and $\mu$, since for larger values our
requirements are more easily fulfilled.

\TABULAR{cc|cccc|cccc}{
& & \multicolumn{4}{c}{$M_1 = 90$ GeV,$\tan\beta = 4$} &
\multicolumn{4}{c}{$M_1 = 120$ GeV,$\tan\beta = 4$} \\
\hline 
$M_2$ & $\mu$ & $ m_{\tilde{\chi}_1^0}$  & $ m_{\tilde{\chi}_2^0}$ &
$ m_{\tilde{\chi}_1^\pm}$  & $|N_{11}|$ & $ m_{\tilde{\chi}_1^0}$  & $ m_{\tilde{\chi}_2^0}$ &
$ m_{\tilde{\chi}_1^\pm}$  & $|N_{11}|$ \\ \hline
 200. &   -600. &    91. &   202. &   202. &   0.88 
&   121. &   202. &   202. &   0.88 \\
 200. &   -200. &    90. &   167. &   170. &   0.85 
&   118. &   168. &   170. &   0.81 \\
 200. &    200. &    77. &   144. &   135. &   0.71 
&    99. &   151. &   135. &   0.56 \\
 200. &    600. &    88. &   191. &   191. &   0.86 
&   118. &   191. &   191. &   0.85 \\
 400. &   -600. &    91. &   396. &   397. &   0.88 
&   121. &   396. &   397. &   0.88 \\
 400. &   -200. &    90. &   197. &   200. &   0.85 
&   118. &   198. &   200. &   0.83 \\
 400. &    200. &    79. &   191. &   181. &   0.80 
&   104. &   196. &   181. &   0.75 \\
 400. &    600. &    88. &   381. &   380. &   0.87 
&   118. &   381. &   380. &   0.87 \\
 600. &   -600. &    91. &   561. &   563. &   0.88 
&   121. &   561. &   563. &   0.88 \\
 600. &   -200. &    90. &   200. &   202. &   0.85 
&   118. &   202. &   202. &   0.84 \\
 600. &    200. &    79. &   201. &   191. &   0.81 
&   105. &   204. &   191. &   0.78 \\
 600. &    600. &    88. &   534. &   533. &   0.87 
&   118. &   534. &   533. &   0.87 \\
\hline 
\multicolumn{10}{c}{$M_1 = 120$ GeV,$\tan\beta = 50$} \\
\hline 
 200. &   -600. &   119. &   197. &   197. &   0.87 & & & & \\
 200. &   -200. &   110. &   158. &   153. &   0.96 & & & &\\
 200. &    200. &   109. &   156. &   150. &   0.67 & & & &\\
 200. &    600. &   119. &   196. &   196. &   0.86 & & & &\\
 400. &   -600. &   119. &   389. &   389. &   0.87 & & & &\\
 400. &   -200. &   112. &   197. &   191. &   0.79 & & & &\\
 400. &    200. &   111. &   197. &   190. &   0.78 & & & &\\
 400. &    600. &   119. &   388. &   388. &   0.87 & & & &\\
 600. &   -600. &   119. &   547. &   547. &   0.87 & & & &\\
 600. &   -200. &   112. &   203. &   197. &   0.81 & & & &\\
 600. &    200. &   111. &   204. &   196. &   0.80 & & & &\\
 600. &    600. &   119. &   545. &   545. &   0.87& & & &\\
}
{Light weak gaugino masses and photino component of 
the lightest neutralino
$|N_{11}|$ for various values of $\tan \beta$, 
$M_1$, $\mu$.  \label{tttt}}

We see that demanding a lightest neutralino in the range
$90-120$ GeV, with the chargino remaining
relatively heavy leads to a neutralino mixing in a photino component
that is significantly constrained, and lies in the range (0.81-0.88).
This also holds for larger values of $M_2,\mu$ which are not included in the table.

We use a single slepton mass parameter 
$m_{\tilde l}\equiv{m_{{\tilde
\mu},{\tilde e}}}_R={m_{\tilde
e,{\tilde \mu}}}_L$.
Neglecting small fermion mass terms, the tree-level slepton masses are then
(for the first two generations) 
\begin{eqnarray}
m_{{\tilde e,{\tilde \mu}}_L} &=& m_{\tilde l}^2 + (\frac{1}{2} - \sin
\theta_W^2) M_Z^2 \cos(2 \beta) \nonumber \\
m_{{\tilde e,{\tilde \mu}}_R} &=& m_{\tilde l}^2 -  \sin
\theta_W^2 M_Z^2 \cos(2 \beta), \nonumber \\
m_{{\tilde \nu}_e,{\tilde \nu}_\mu} &=& m_{\tilde l}^2 +  \frac{1}{2}\sin
\theta_W^2 M_Z^2 \cos(2 \beta)
\label{slepmasses}
\end{eqnarray}
with negligible mixing proportional to the electron and muon masses
respectively. 

We use the {\small \tt ISASUSY} part of the {\small \tt ISAJET7.58}
package~\cite{isajet} to generate the spectrum, branching ratios and 
decays of the sparticles. 
For an example of parameters, we choose (in the notation used by
ref.~\cite{isajet}) $\lambda'_{211}=0.01$, $m_{3/2}=10^{-3}$ eV, $\tan
\beta=10$, $A_{t,\tau,b}=0$, 
$\mu$ together with 
other flavour diagonal soft supersymmetry breaking parameters are set to
2000 GeV. We emphasise that this is a representative point in the
supersymmetric parameter space and not a special choice. 
Any superparticles except the first two
generation 
sleptons, the lightest neutralino and the gravitino do not appear in this
analysis because they are too heavy to be produced or to contribute to cascade
decays in CDF data.

\section{Fitting kinematical distributions}

We now simulate the signal events for the process in Figure~\ref{feyn}.
The Standard Model background is taken from ref.~\cite{CDF}.
We use {\small \tt HERWIG6.3}~\cite{herwig} including parton showering (but
not including isolation cuts) to calculate cross-sections for single slepton
production. The slepton mass parameter 
$m_{\tilde l}$
 and the bino mass parameter\footnote{All
parameters take their quoted values at the electroweak scale.} $M_1$ are
allowed to vary in order to see what range of neutralino and slepton masses
are preferred by the experiment.

We simulate the detector by the following:
\begin{itemize}
\item
Photons can be detected if they do {\em
not} 
have rapidity $1.0<|\eta|<1.1$, $|\eta|<0.05$.
The region $0.77<\eta<1.0, 75^\circ<\phi<90^\circ$ is also excluded because it
is not instrumented. If these constraints are satisfied, we assume 81$\%$
detection efficiency for the photons.
\item
Muons have a 60$\%$ detection efficiency if $|\eta_\mu|<0.6$ or 45$\%$
if $0.6\leq \eta_\mu\leq 1.1$
\end{itemize}
Rapidity of a particle is defined as $\eta= - \ln \tan (\theta/2)$, where
$\theta$ is the longitudinal angle between the particle's momentum and 
the beam. $\phi$ is the transverse angle between the particle's momentum and
the $x$-axis. We also implement the cuts used in the experimental analysis to
beat the background down:
$E_T(\mu)>25$ GeV, $E_T(\gamma)>25$ GeV and $\mett>25$ GeV. Because we do not
perform jet reconstruction, we do not perform isolation cuts.

CDF gave one-dimensional projections in the $\mu \gamma \mett$ events for the
following kinematic variables
\begin{eqnarray}
E_T = \sqrt{p_x^2 + p_y^2}, \nonumber \\
m_{12} = \sqrt{p_1.p_2} \nonumber\\
M_T^2 = E_T^2 - p_x^2 - p_y^2\nonumber\\
\Delta \phi_{12} = \phi_1 - \phi_2  \nonumber \\
H_T = \mett + E_T(\gamma) + E_T(\mu) \nonumber\\
\Delta R_{\mu \gamma} = \sqrt{\Delta \phi_{\mu \gamma}^2 + (\eta_\mu-\eta_\gamma)^2}
\end{eqnarray}
where $p_{x,y}$ are the $x$ and $y$ (i.e. transverse) components of the
momentum respectively and $p_{1,2}$ refer to 4-momenta of particles labeled
by 1 and 2. 

Once we have a statistically large sample of signal events simulated, we
have a prediction for the number of expected signal plus background events in
bin $i$ of a distribution: $N_{S+B}^i$. 
We use the average background presented in ref.~\cite{CDF}.
The Poisson
distributed probability density function (PDF) of observing $N_{obs}^i$ events
compared to $N_{S+B}^i$ is
\beq
p^i_{S+B}(N^i_{obs},N^i_{S+B}) = \frac{e^{-N^i_{S+B}}
{(N^i_{S+B})}^{N^i_{obs}}}{r}. 
\eeq
In any one distribution, $p_{S+B} \equiv \Pi_i p^i_{S+B}$ gives the total PDF
for that 
distribution, assuming each bin is uncorrelated to the others.
Unfortunately, we are not able to make this assumption between
different distributions of variables, because they contain data on the same
events and should contain some level of correlation. 
Finally, the log likelihood of signal plus background 
$\ln p_{S+B}$  is calculated 
and normalised by subtracting the analogous log
likelihood 
for the Standard Model background prediction
\beq
-2  \ln L \equiv -2(\ln p_{S+B} - \ln p_{B}). \label{logL}
\eeq
A negative value then indicates that the data favour the signal plus
background hypothesis over just background.
When performing parameter ($x_i$) estimation, 
one determines the equivalent number of
$\sigma$ 
away from the best-fit point (which has parameters $x_b$) by
\beq
(\Delta\sigma)^2 = -2 \left( \ln L(x_i) - \ln L(x_b) \right). \label{numsig}
\eeq
The value of $\ln L(x_i)$
calculated is then equivalent to a probability which matches the number of
$\Delta\sigma$ that our model fits the data better than the Standard Model in 
conventional Gaussian statistics. Thus, the measure of number of
$\Delta\sigma$ here 
is purely a measure of probability in tails of PDFs, not a statement about the
Gaussian nature of that PDF.

We now perform a fit to $M_1$ and $m_{\tilde l}$ keeping all other fundamental
parameters (but not branching ratios) constant.
We must perform the fit one time for each different distribution; 
because the distributions come from the same events, they are all correlated
to some extent. We cannot therefore assume that the distributions are
uncorrelated in order to fit more than one distribution at a time, i.e. the
correlations must be taken into account.
While we can generate the correlations in the signal events by
Monte-Carlo, we do not own multi-dimensional data on the kinematic variables
in the 
data. We therefore cannot perform a fit to more than one distribution at any
time.  
However, it is possible to see to what extent the individual fits to each
variable are 
compatible with each other.
The fit corresponds to maximising the log likelihood obtained from
Eq.~\ref{logL} for one of the distributions.
In Table~\ref{bestfit}  we present the best fit points, indicating that
light neutralino masses,  in the range of 67-111 GeV 
are to be expected.
For small $M_1$ the lightest neutralino mass is determined by
$M_1$, while $M_2$ controls the chargino mass. If $M_2$ is heavy (as assumed
here), 
no cascade decays involving charginos are kinematically favoured. 
\TABULAR{c|cccc}
{\label{bestfit}
Variable  &   $M_1$ (GeV) &    $m_{\tilde{\ell}}-M_1$ (GeV) &  $-2\Delta\ln(L)$ &
$\Delta\sigma$ \\ \hline
$E_T(\mu)$  &   87   &   35   &  -10.94 & 3.31\\
$E_T(\gamma)$  &   67  &   30   &  -9.36 & 3.06 \\
$\mett$  &   104   &   47   &  -6.09 &  2.47\\
$m_{\mu \gamma}$  &   82   &   23    &  -9.94 & 3.15  \\
$M_T(\mett\mu)$  &   96    &   44    &  -6.54 & 2.56   \\
$M_T(\mett\gamma)$  &   99   &   24   &  -8.81 &  2.97\\
$M_T( \mett\gamma \mu)$  &   84   &   28  &  -6.30 &  2.51 \\
$\Delta \phi_{\mett \gamma }$  &    111  &   33   &  -5.28 & 2.30 \\
$\Delta \phi_{\mu \gamma}$   &   97  &   26   &  -8.87 & 2.98   \\
$\Delta \phi_{\mett \mu}$  &   99  &   25  &  -9.01 & 3.00  \\
$H_T$ & 72 & 33 & -5.18 & 2.28\\
$\Delta R_{\mu \gamma}$ & 83 & 24 & -6.26 & 2.50 \\
}{Separate best fit points for each kinematic variable. We display the bino
mass 
parameter $M_1$, the mass splitting between $M_1$ and the slepton mass
parameter $m_{\tilde l}$. We display the difference in log likelihood between
our model 
and the Standard Model $-2\Delta \ln (L)$ and the corresponding number of
sigma $\Delta \sigma$ that the model fits the kinematic distribution better
than the Standard Model.}

\TABULAR{c|c}
{\label{tab:res1}
cut & percentage \\ \hline
detected $\mu$ & 52.1 \\
$\mett>$  25 GeV & 41.7\\
detected $E_T(\gamma)>25$ GeV & 20.8\\
detected $E_T(\mu)>25$ GeV & 11.4\\ \hline
$\sigma$ & 0.091 pb \\
}{Percentage of SUSY events that satisfy cumulative cuts for $\mu \gamma
\mett$ events at CDF, Run I for the best-fit point. Events that pass a cut on
a given row also pass 
those cuts on rows above.
The $\gamma$-in-active-region cut is described
in the text. The cross-section after all cuts is displayed on the last row.} 
For the rest of this section, we concentrate on the best-fit point for $E_T(\mu)$, because
this gives the best likelihood out of all the fits. 
In Table~\ref{tab:res1}, we show the percentage of events making it through
each of the cuts for this best-fit point. 
The table shows that 11.4$\%$ of sleptons produced end up as detected
$\mu \gamma \mett$ events in CDF. The cross-section of 0.091 pb predicts 
a total of 7.86 events in the $\mu \gamma \mett$ channel. This is higher than
the observed excess because the $E_T(\mu)$ distribution itself prefers it.

The relevant branching ratios of the smuon for this point are
\begin{eqnarray}
BR({\tilde \mu}_L \rightarrow \chi_1^0 \mu) = 0.984, \nonumber \\
BR({\tilde \mu}_L \rightarrow \bar{u} d) = 0.015, \nonumber \\
BR({\tilde \mu}_L \rightarrow {\tilde \mu} {\tilde G}) = 0.001,
\end{eqnarray}
with a lifetime of $1\times 10^{-23}$ sec,
whereas for the lightest neutralino we have
\begin{eqnarray}
BR(\chi^0_1 \rightarrow {\tilde G} \gamma) = 0.975, \nonumber \\
BR(\chi_1^0 \rightarrow {\tilde G} e^- e^+) = 0.019,
\end{eqnarray}
with a lifetime of $1 \times 10^{-19}$ sec. At such small values of
$\lambda'_{211}$ and $m_{\tilde G}$, R-parity violating decays of the 
lightest neutralino are negligible. The light sparticle masses are
\beq
m_{\chi_1^0} = 86.8\mbox{~GeV}, \ 
m_{{\tilde e}_L, {\tilde \mu}_L} = 130.8\mbox{~GeV}, \
m_{{\tilde \nu}_L}=104.2 \mbox{~GeV}, \
m_{{\tilde e}_R, {\tilde \mu}_R} = 129.7\mbox{~GeV},
\eeq
whereas we have set all of the other sparticles except for the gravitino to be
heavy (around 2000 GeV), 
so that they play no role in our analysis.

\FIGURE{%
\fourgraphs{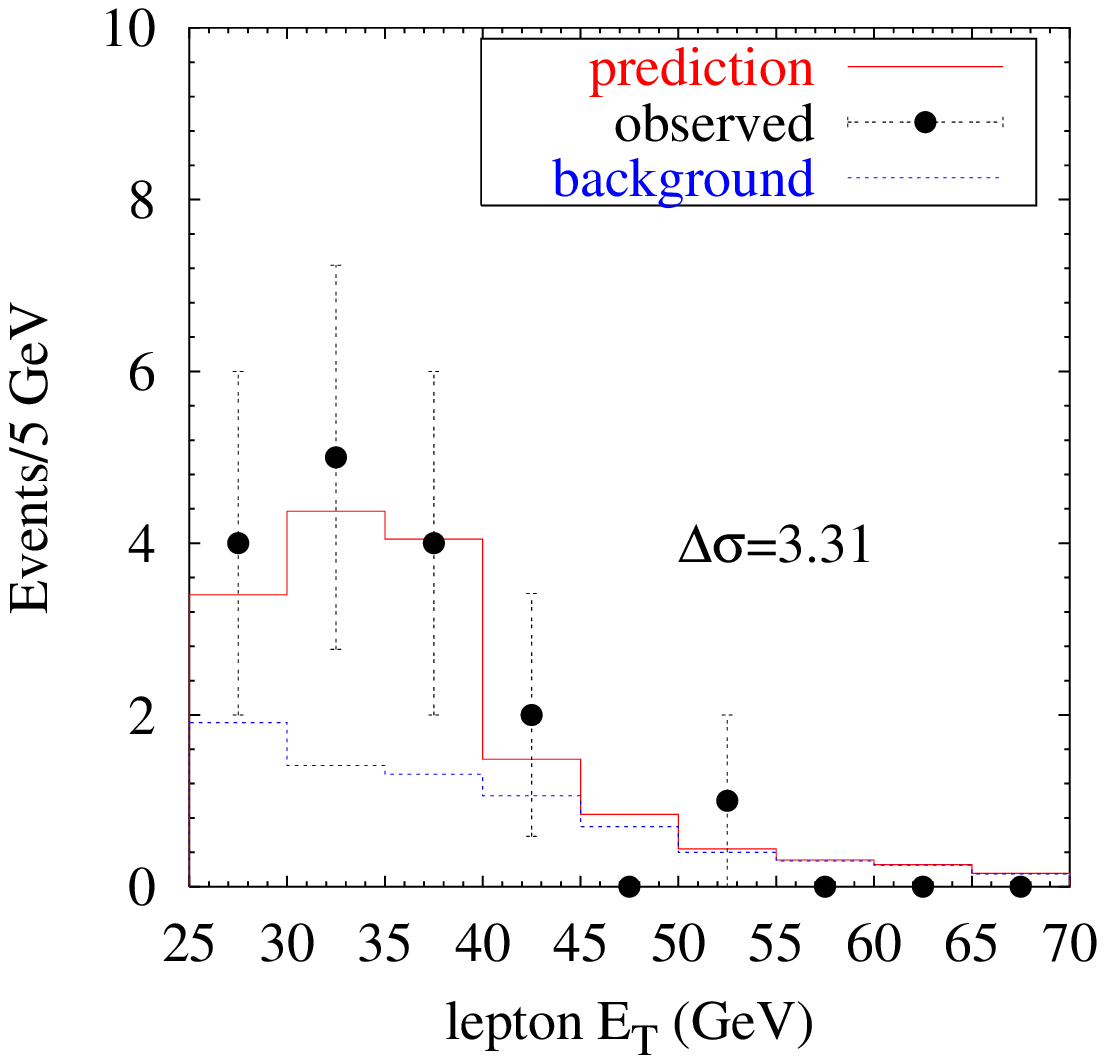}{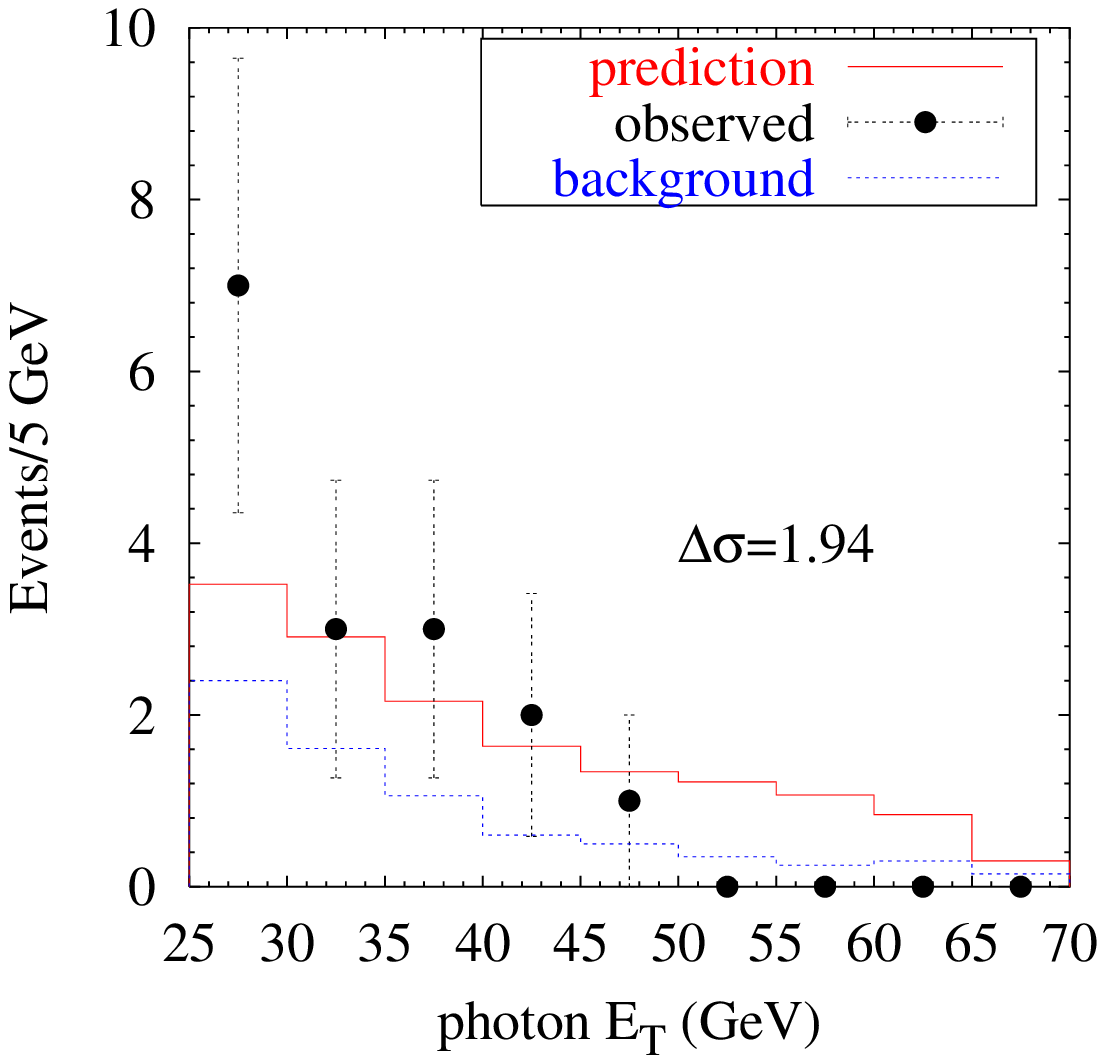}{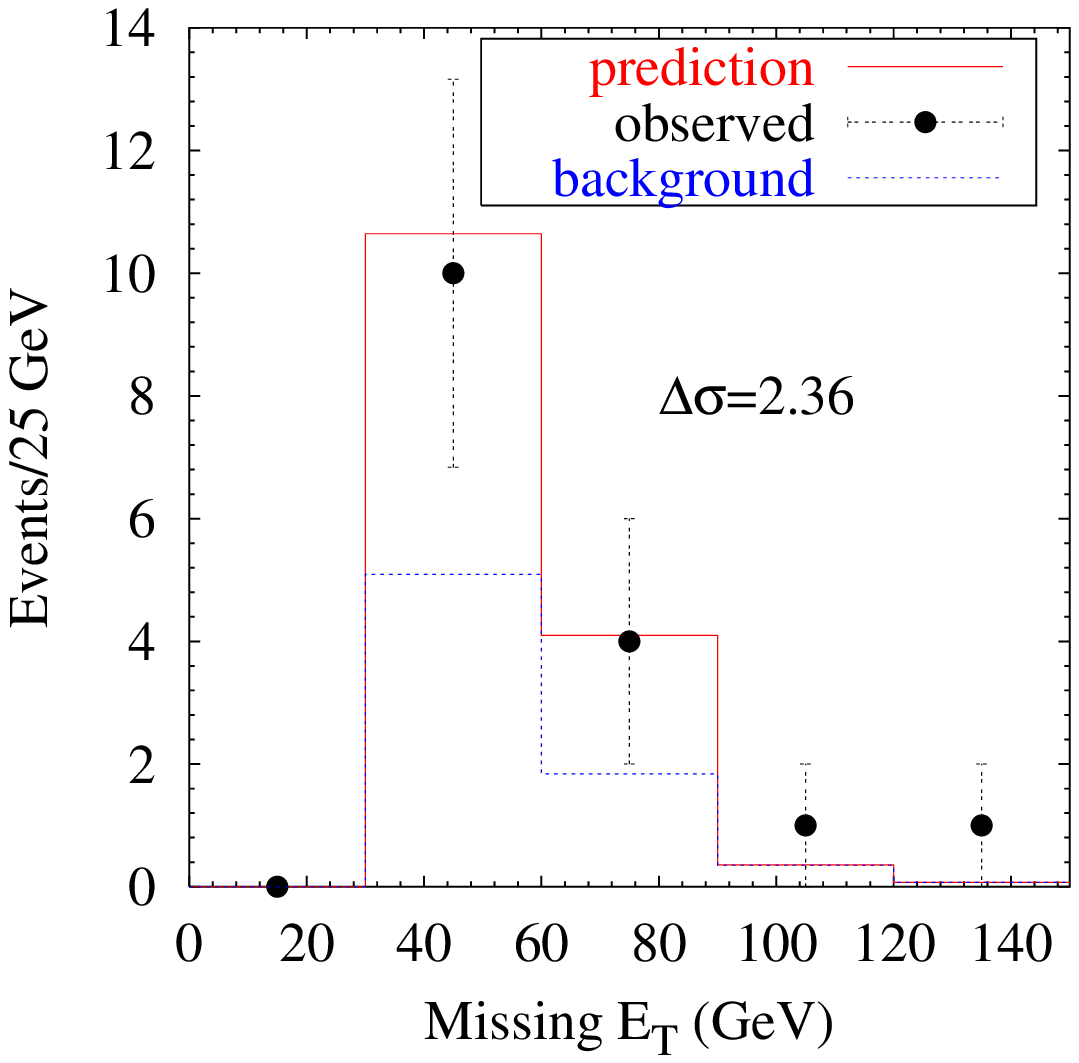}{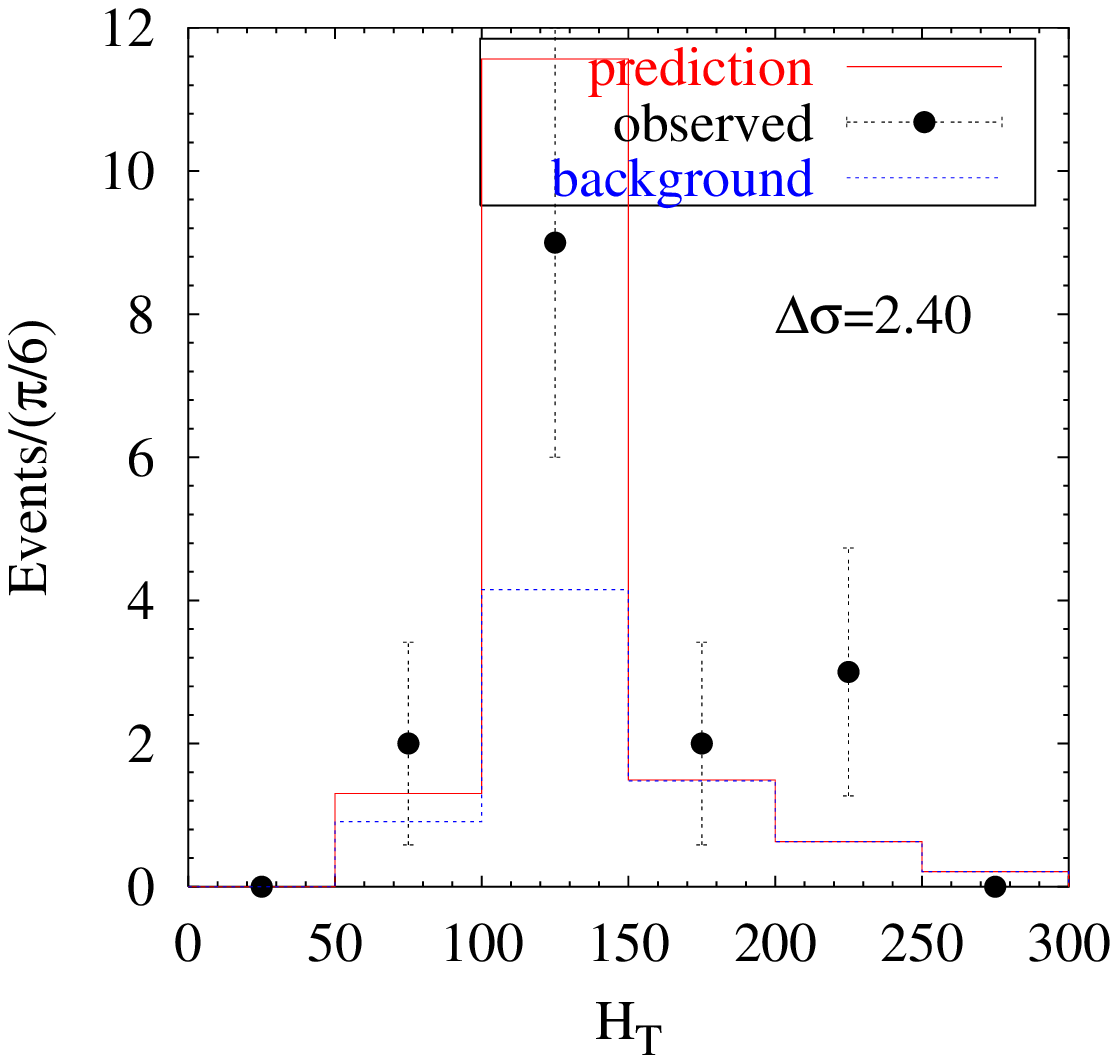}
\label{dists}
\caption{Energy distributions for the $\mu \gamma \mett$ events.
We show the distributions in  (a) lepton $E_T$, (b)
photon $E_T$, (c) $\mett$ and (d) $H_T$.
The solid red histogram is signal plus
background for our best-fit point, the blue dashed histogram is the Standard Model background 
and the black points (with $\sqrt{N}$
error-bars imposed) are the observed number of events. 
}}
\FIGURE{%
\fourgraphs{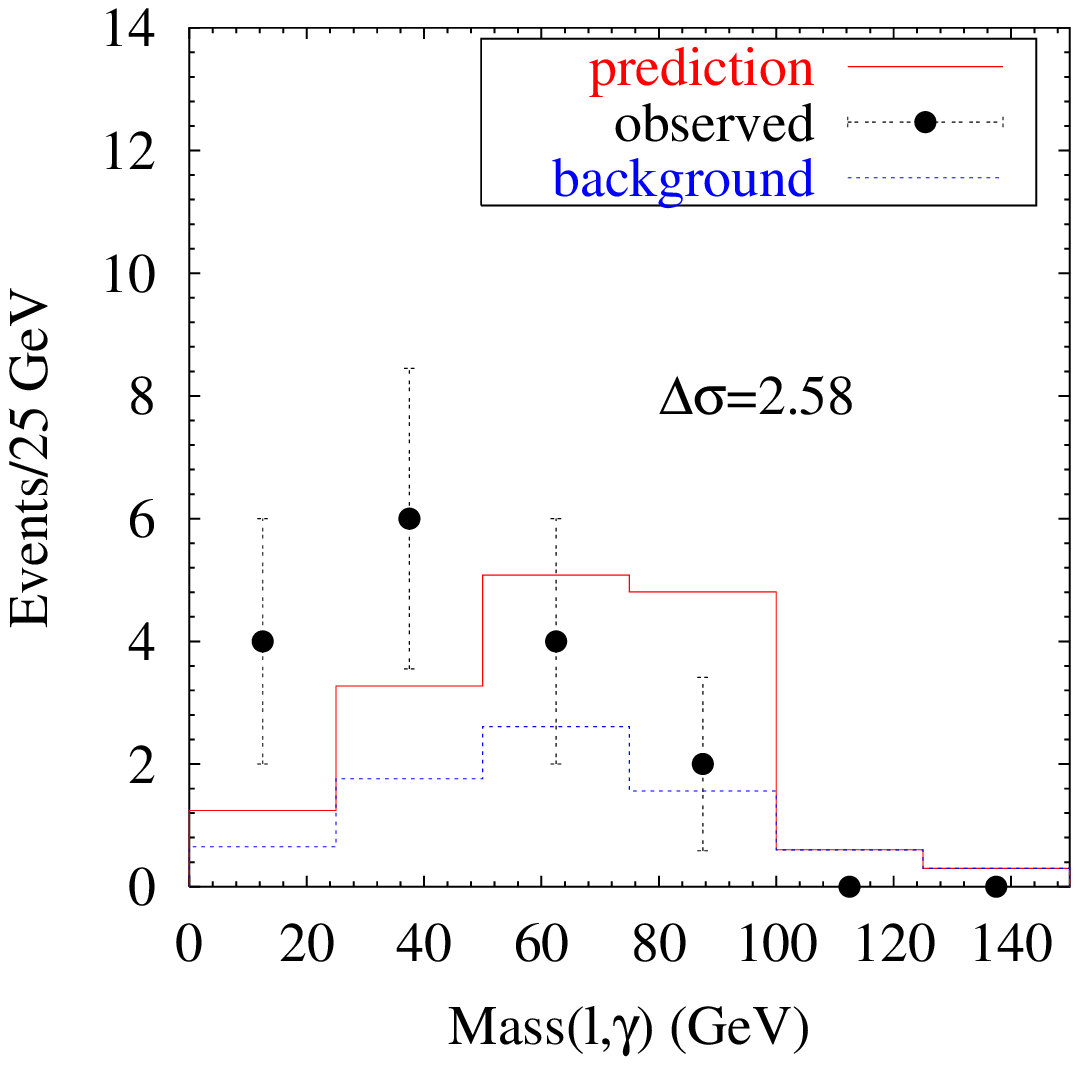}{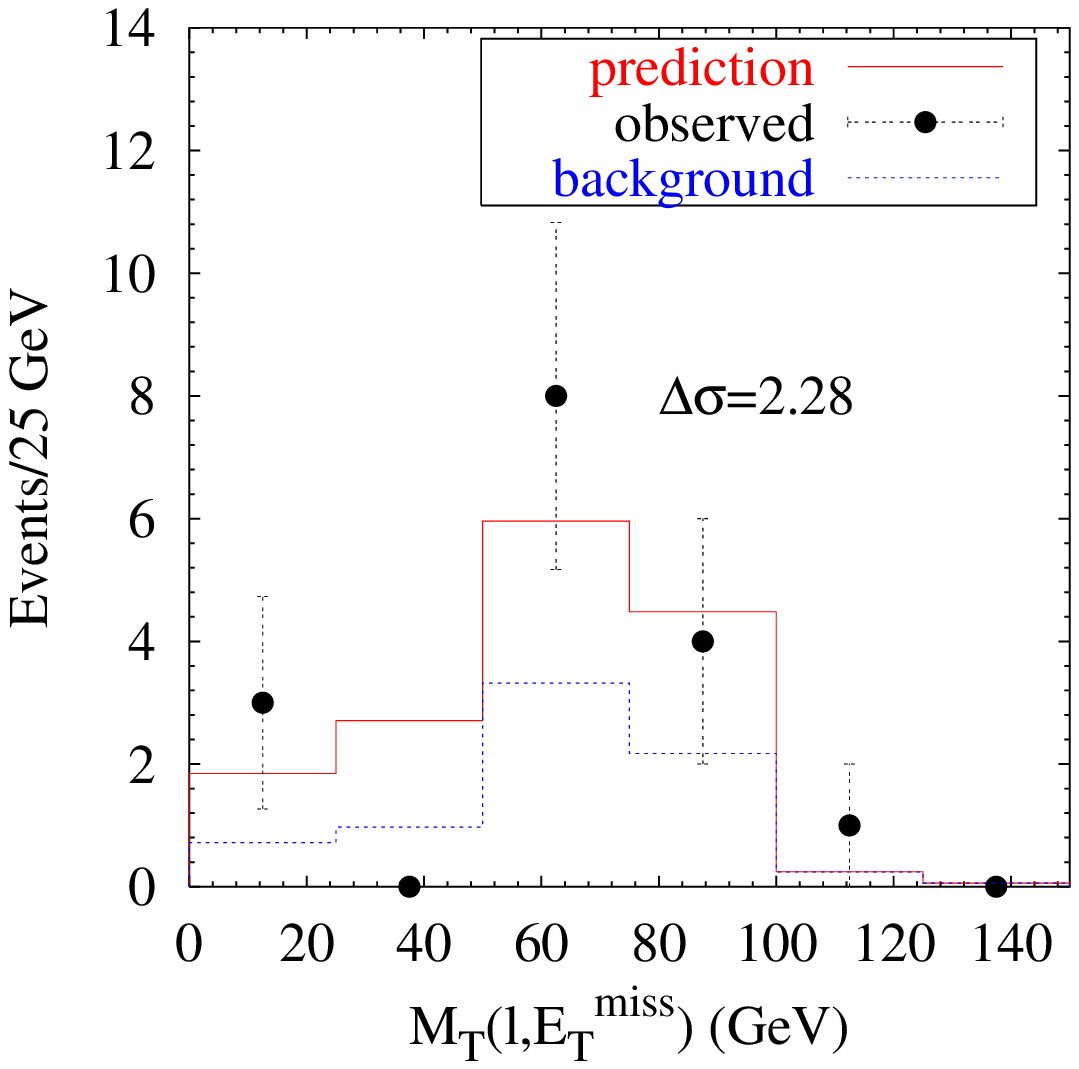}{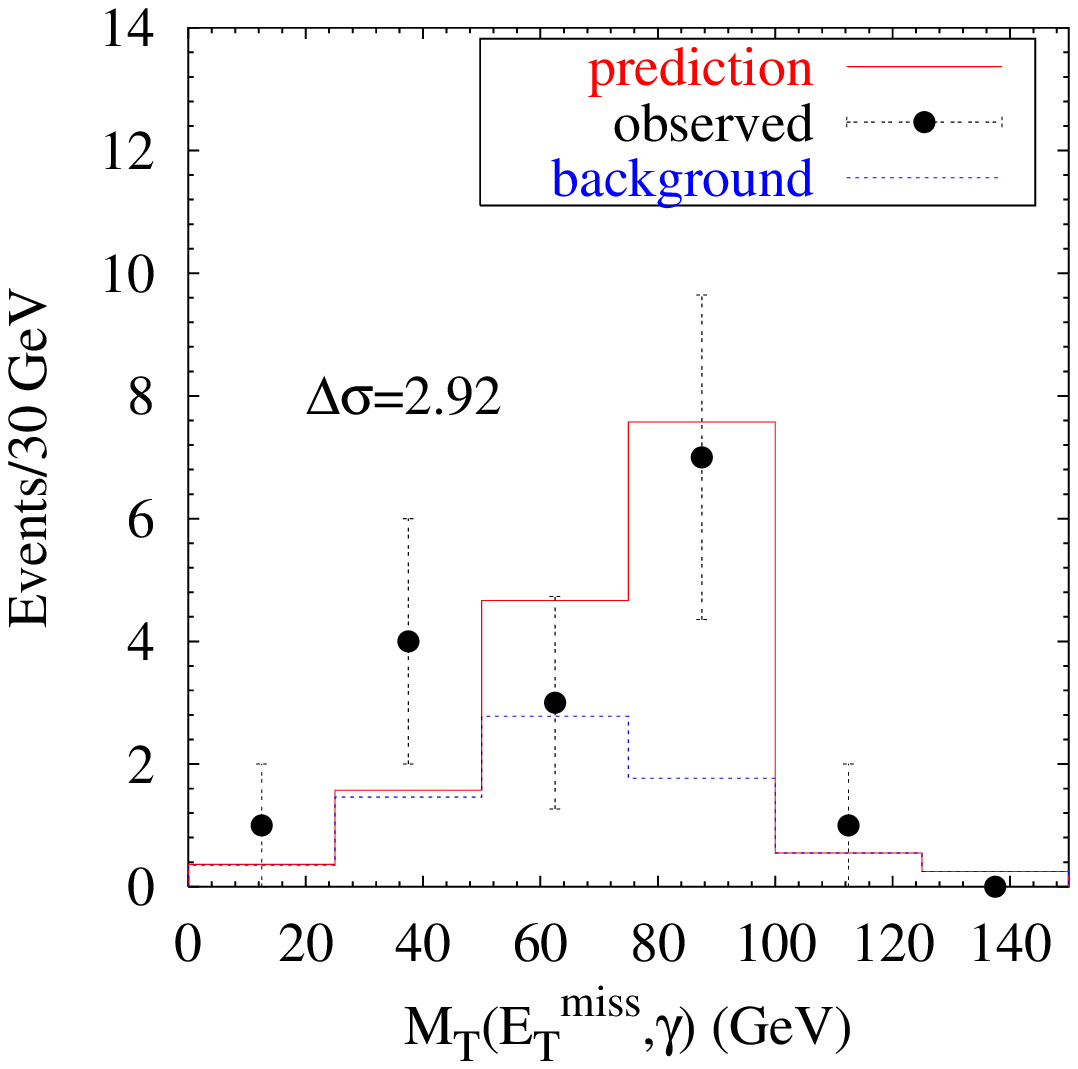}{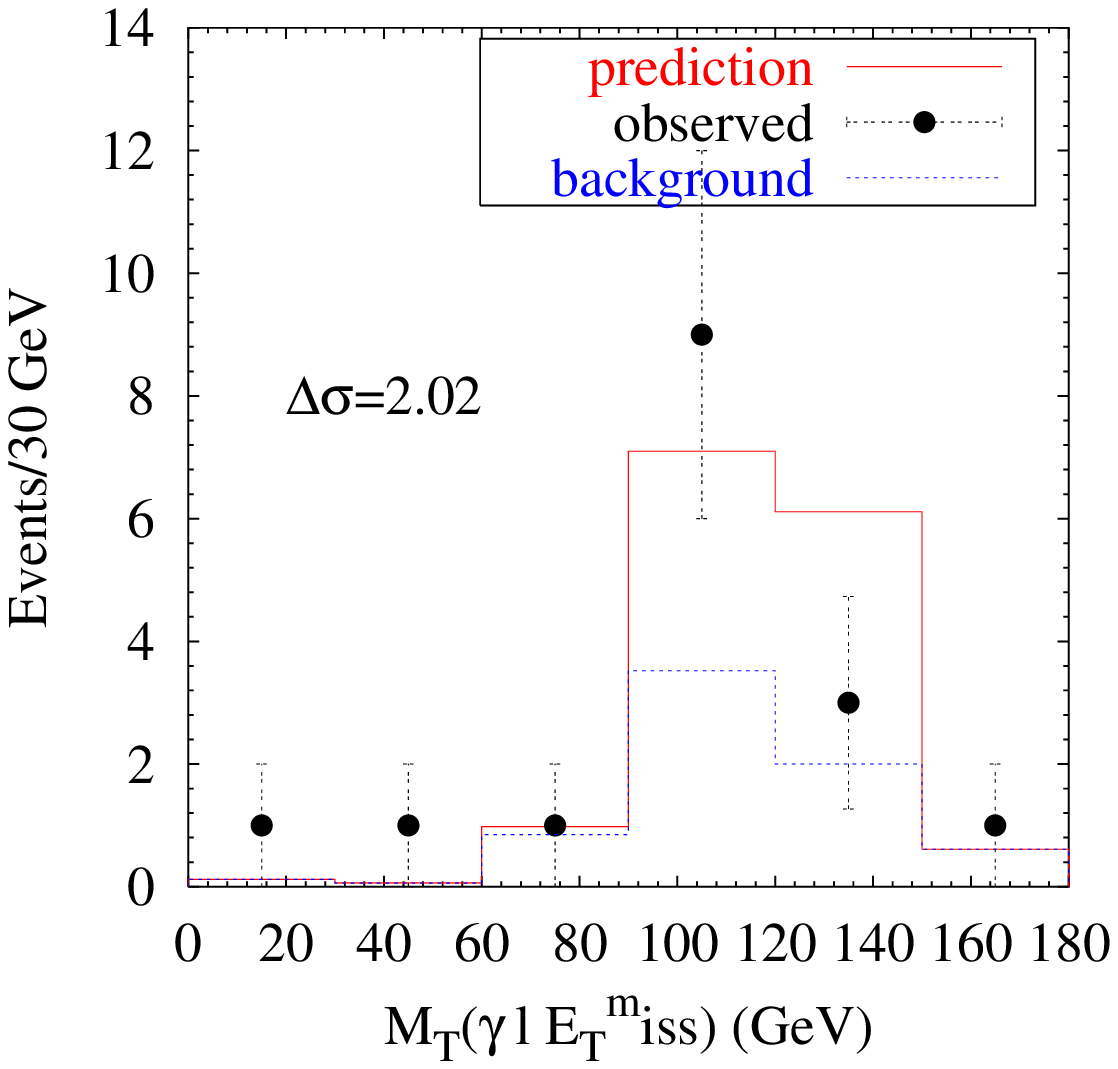}
\label{dists2}
\caption{Mass distributions for the $\mu \gamma \mett$ events.
We show the distributions in (a) the invariant mass of the
$\mu$-$\gamma$ pair, (b)
transverse mass of $\mu \mett$, (c) transverse mass of
$\gamma-\mett$ and (d) transverse mass of the $\gamma-\mett-\mu$ subsystem.
The solid red histogram is signal plus
background for our best-fit point, the blue dashed histogram is the Standard Model background 
and the black points (with $\sqrt{N}$
errorbars imposed) are the observed number of events.
}
}
\FIGURE{%
\fourgraphs{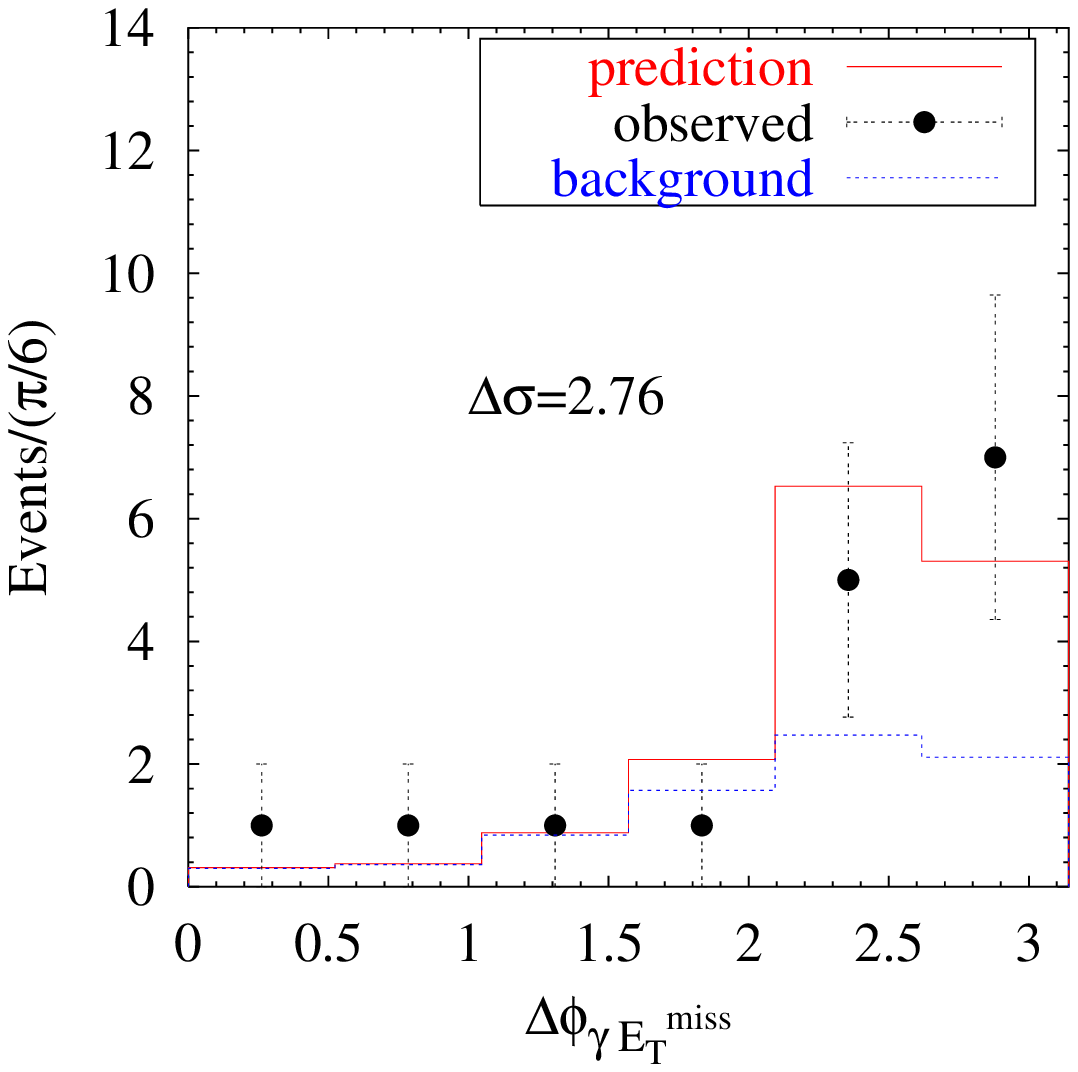}{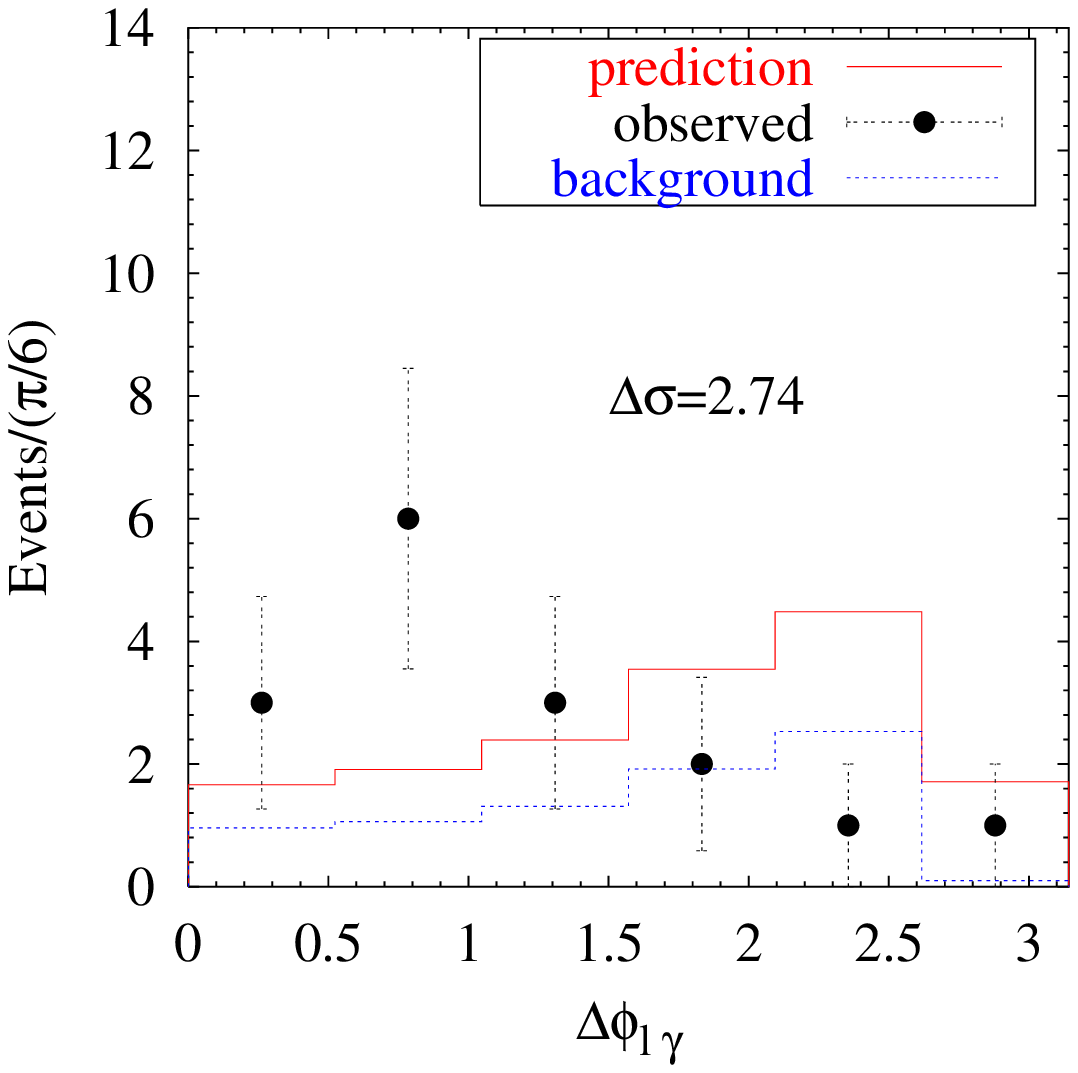}{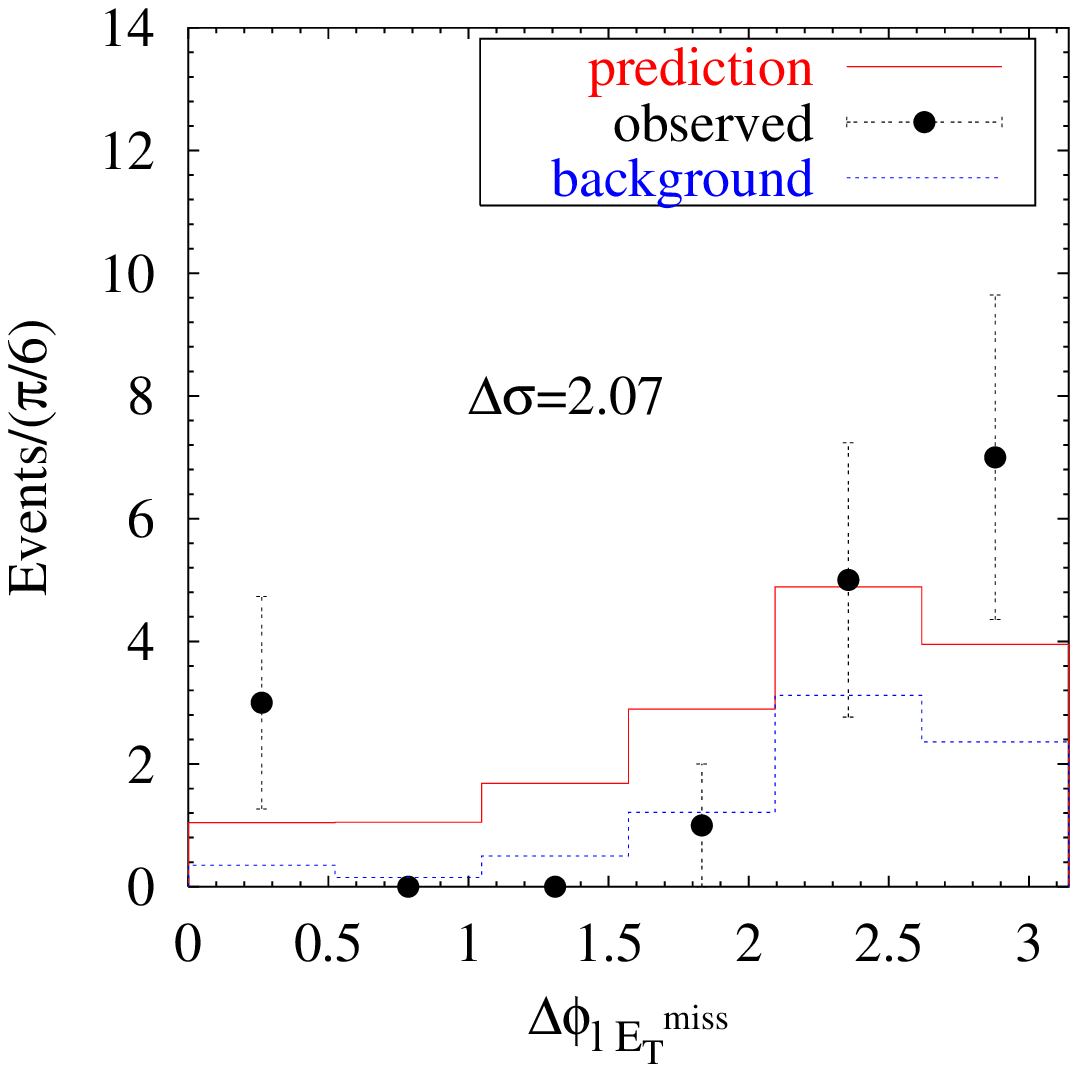}{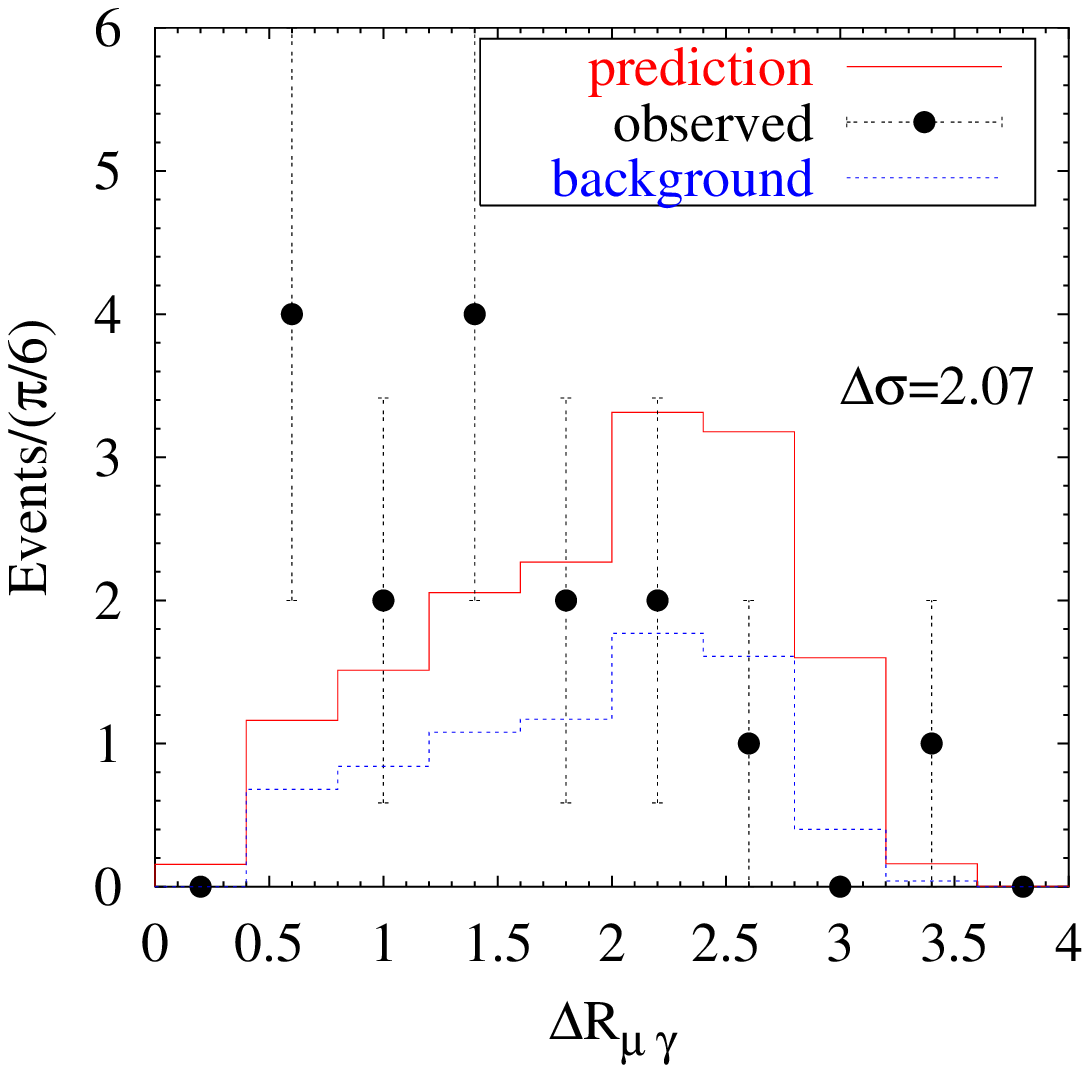}
\label{dists3}
\caption{Transverse angular distributions for  $\mu \gamma \mett$ events.
We show the distributions in  the transverse angle between (a) $\gamma \mett$,
(b) $\mu \gamma$, (c)  $\mu \mett$ and (d) $\Delta R_{\mu \gamma}$.
The solid red histogram is signal plus
background for our best-fit point, the blue dashed histogram is the Standard Model background 
and the black points (with $\sqrt{N}$
errorbars imposed) are the observed number of events.
}
}
For this best-fit parameter point, we show the predicted distributions
of lepton $E_T$, photon $E_T$ and $\mett$ in the histograms of
Fig.~\ref{dists} and compare them with the excess of the data over the
Standard Model background. 
$\Delta \sigma$ is
labeled on each plot and is the equivalent number of sigma that this best-fit
point fits a particular distribution better than the Standard Model.
Obviously the largest $\Delta \sigma=3.31$ is for the lepton $E_T$, since the
fit is performed to this variable. But we also see that at this point, the
other distributions also fit the data better than the Standard Model: all fit
the data better than the Standard model to, at least, 2$\sigma$ except for
the fits to $E_T(\gamma)$ for which $\Delta \sigma=1.94$. The photon $E_T$ 
seems to be steeper in
the data than in either the Standard Model or in our model and this is a 
feature at other values of $M_1, m_{\tilde l}$.
In Fig.~\ref{dists2} we show the mass distributions. The data seems to
indicate a bump extra to the Standard Model at lower values of $m_{\mu
\gamma}$, as shown in  Fig.~\ref{dists2}a. The angular distributions in
Fig.~\ref{dists3} show that our best-fit point fits the observed excess well in
events where 
the $\gamma$ and $\mett$ are roughly back-to-back in Fig.~\ref{dists3}a. While
Fig.~\ref{dists3}b does not seem a particularly better fit than the Standard
Model by eye, nearly all of the difference in $\ln L$ comes from the last bin,
where the Standard Model predicted hardly any events.

\FIGURE{
\vtwographs{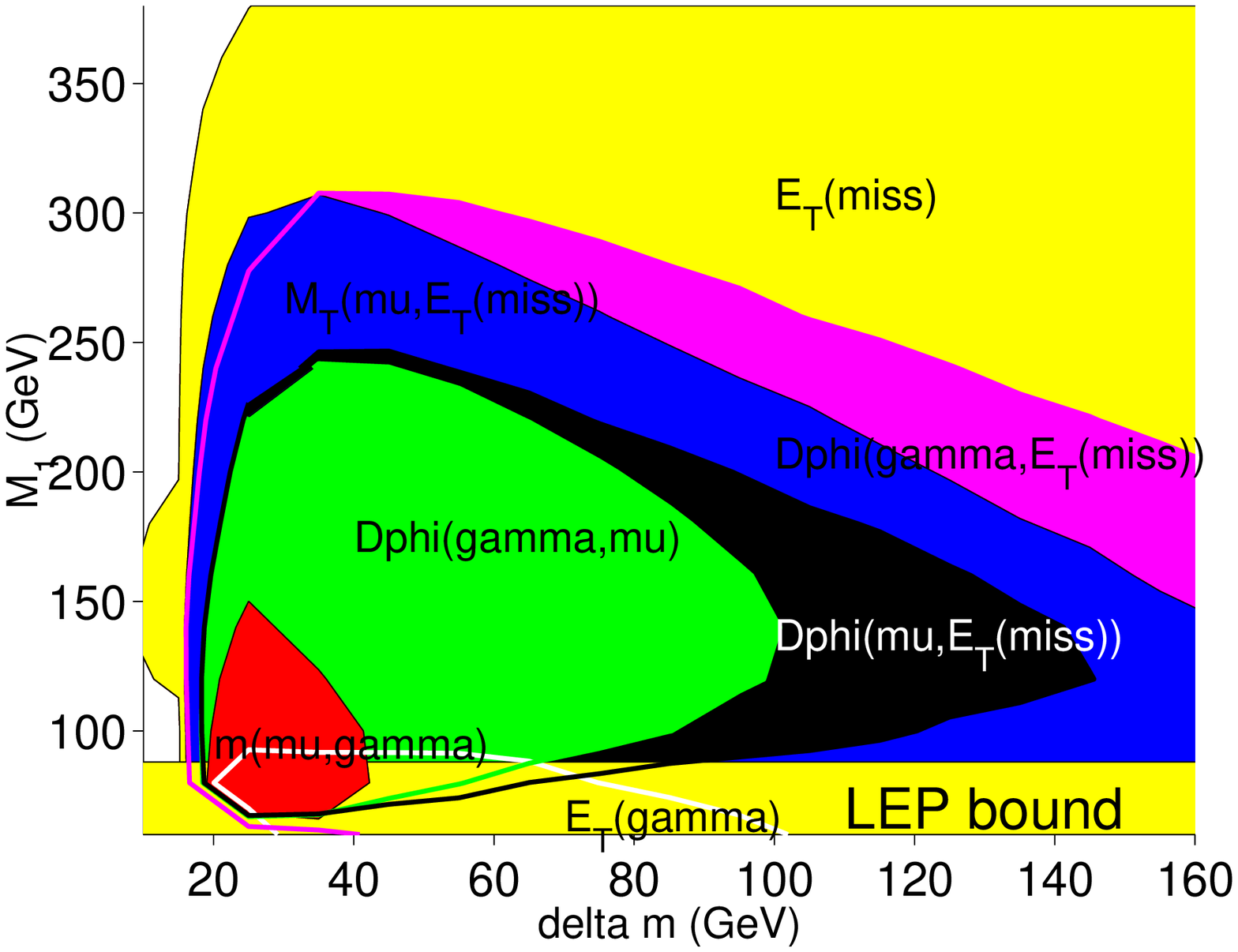}{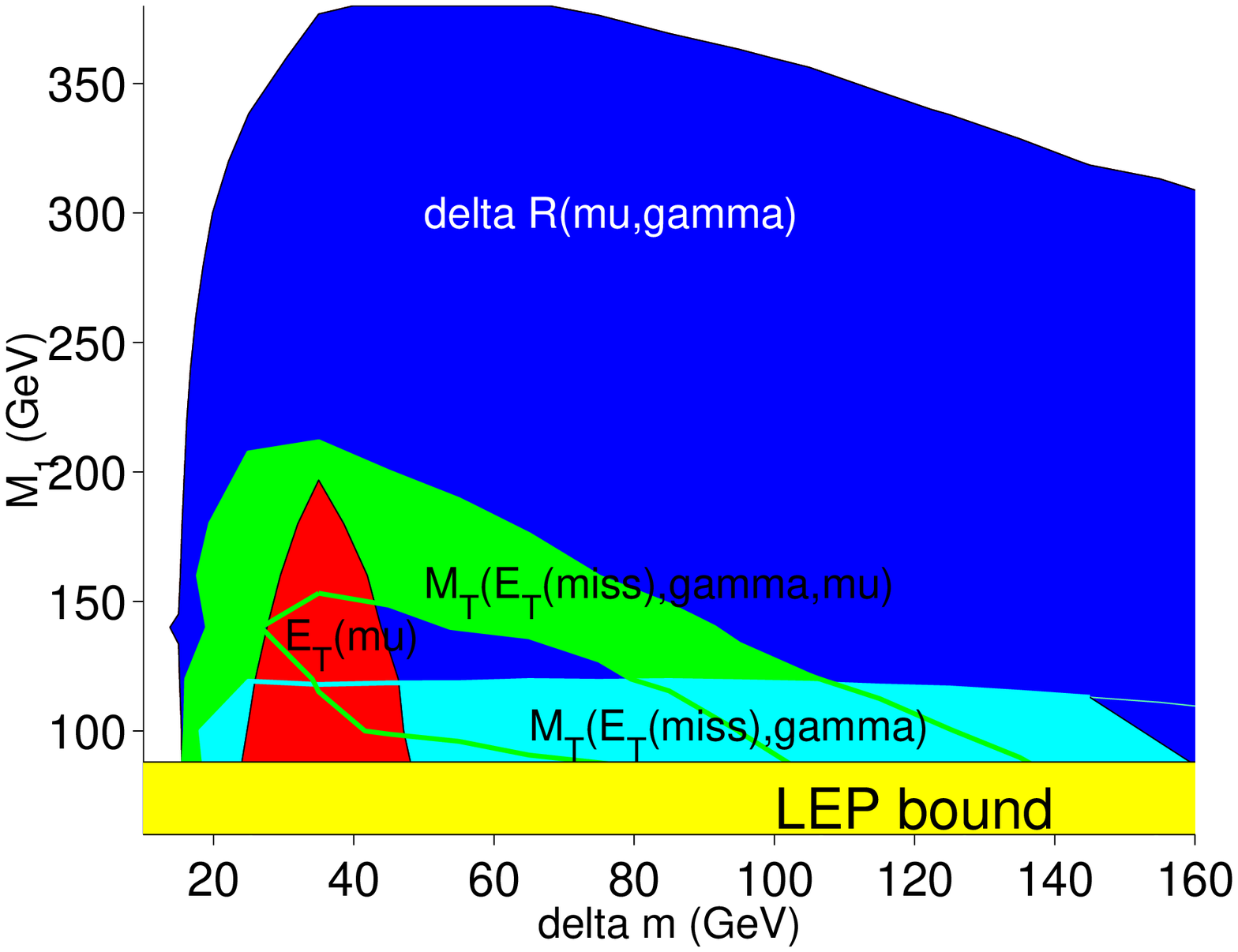}
\label{scans}
\caption{Separate good fit regions for each kinematical distribution. The horizontal
region at the bottom of 
the plot displays the LEP bound from neutralino pair production where the
neutralinos decay to photons and $\mett$~\cite{PDG}. 
The horizontal axis is $\Delta m=m_{\tilde
l - M_1}$. The kinematical variables are (a) $\mett$ (yellow), $\Delta
\phi_{\gamma \mett}$ (magenta), $M_T(\mu\mett)$ (blue), $\Delta \phi_{\mu
\mett}$ (black), $\Delta \phi_{\gamma \mu}$ (green), $m_{\mu \gamma}$ (red),
$E_T(\gamma)$ (white line)
and
(b) $E_T(\mu)$ (red), $M_T(\mett, \gamma, \mu)$ (green), $M_T(\mett,
\gamma)$ (cyan) and $\Delta R_{\mu \gamma}$ (blue).
}
}
To calculate 95$\%$ $C.L.$ regions, we scan over the parameters $\Delta m\equiv
m_{\tilde l}-M_1$ and $M_1$, calculating $(\sigma)^2$ from eq.~\ref{numsig}
at each point and for each kinematical distribution. The 95$\% C.L.$ is then
given by $(\sigma)^2=-2 \Delta \ln L_{BF}+5.99$, where $\ln L_{BF}$ is the 
log likelihood at the best-fit point of the kinematic variable being examined.
The 95$\%$ $C.L.$ regions of $\Delta m$ and $M_1$ for each
separate fitted kinematical distribution are displayed in Fig.~\ref{scans}.
The horizontal
region at the bottom of 
the plot displays the LEP bound from neutralino pair production where the
neutralinos decay to photons and $\mett$~\cite{PDG}. We
note~\cite{hutchmonkey} that analysis of LEP2 data at the highest energies
should be able to cover the region up to $M_1=100$ GeV or so. The ``overlap'' region
$\Delta m\approx 30-40$ GeV and $M_1\approx 90-120$ is encouragingly within
the 95$\%$ confidence-level regions for all distributions except for 
$E_T(\gamma)$ (shown as the area inside the white line in Fig.~\ref{scans}a),
which prefers $M_1<90$ GeV, below the LEP bound.  
Ideally, a
correlated fit would be performed to all distributions simultaneously. 
Then, the significance of not having such a good fit for $E_T(\gamma)$ 
in the overlap region could be calculated.
The most constraining variables are $E_T(\mu)$ and $m_{\mu \gamma}$, which
require $(\Delta m,M_1)<(50,200)$ GeV and $(40,150)$ GeV 
respectively. The $M_T(\mett,\gamma)$ region constrains $M_{\chi_1^0}$ 
to be less than 120 GeV.
 The $H_T$ variable is not plotted because it does not constrain
any of the parameter space at the 95$\%$ C.L.
\FIGURE{
\twographs{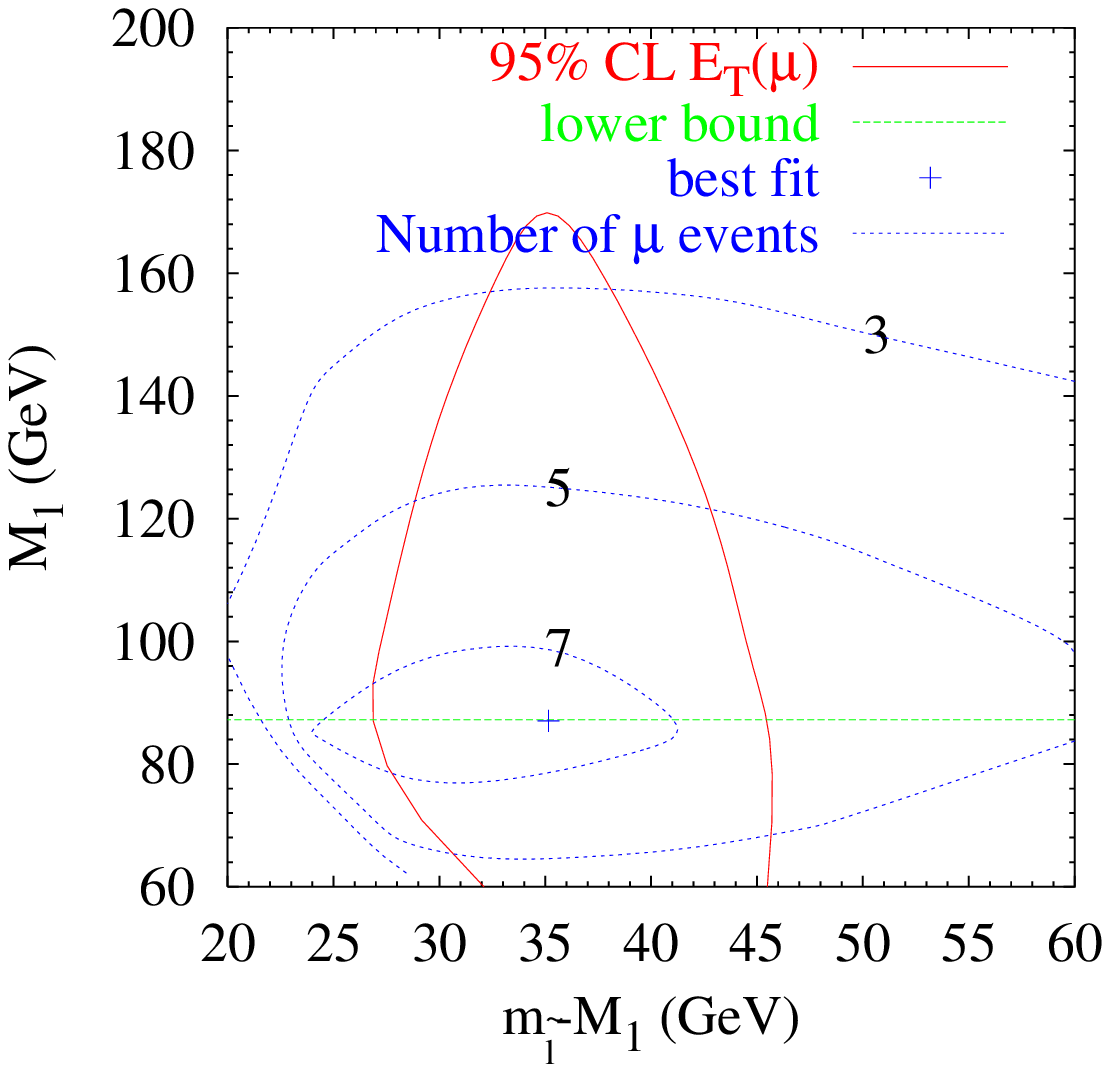}{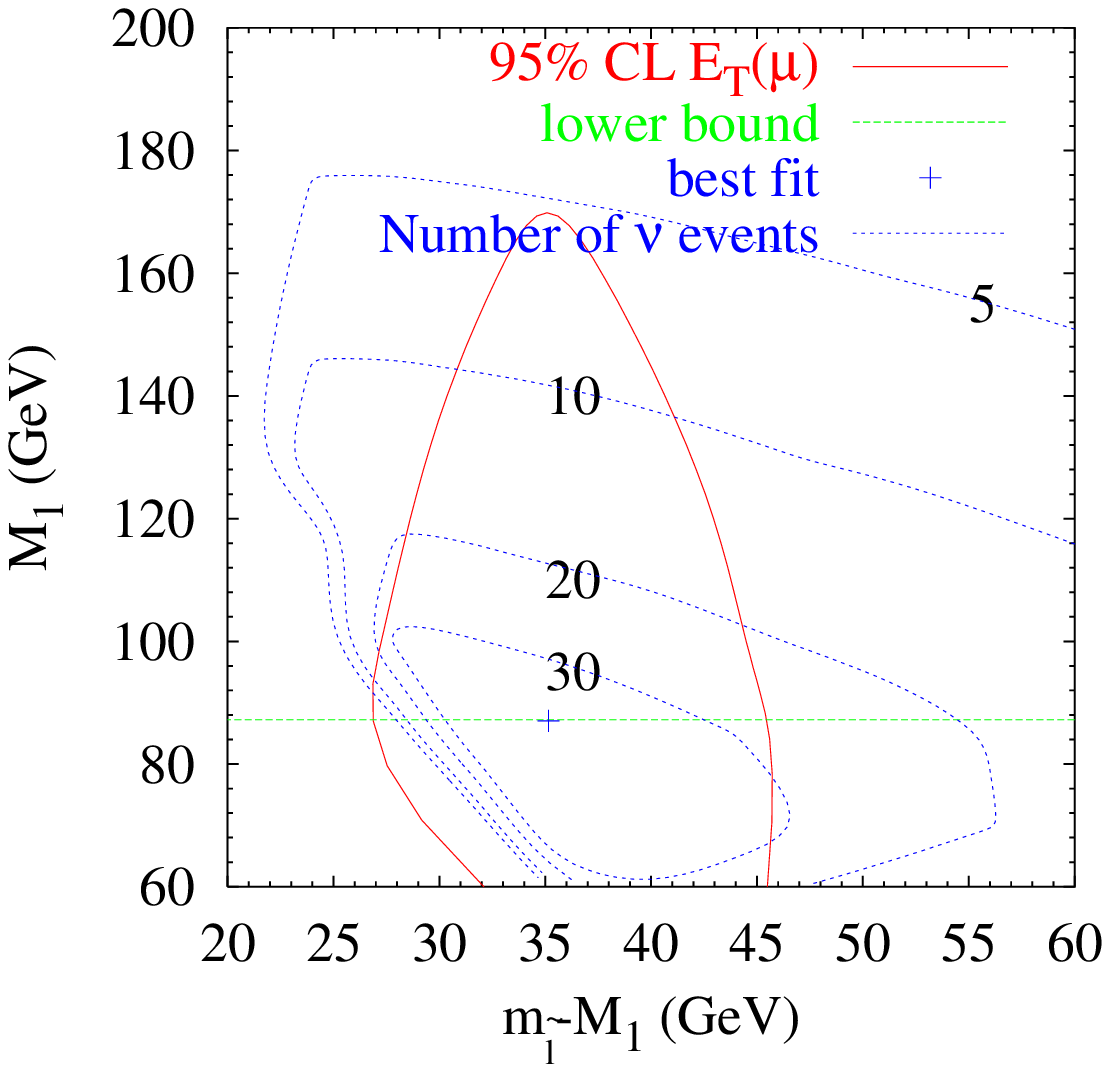}
\label{runIexpected}
\caption{Expected number of (a) $\mu \gamma \mett$ and (b) $\gamma \mett$
signal events in Run I data.
The dashed blue curves show (labeled) contours of number of expected events,
the dotted green
line shows the lower bound coming from LEP2 and the red curve shows the 
region of good fit $E_T(\mu)$ in the for $\mu\gamma\mett$ events.}
}

We now display predictions for various quantities overlaid upon the 95$\%$
C.L. region from the $E_T(\mu)$ distribution for different values of $M_1$ and
$\Delta m$. For example, in Fig.~\ref{runIexpected}a, it is shown that the
expected number of detected signal $\mu \gamma \mett$ events (including the
cuts 
described above) is 3-7 in the 95$\%$ C.L. region. 
Each parameter point specifies a particular sneutrino mass by eq.~\ref{slepmasses}, and
the $\lambda'_{211}$ coupling will also lead to resonant production of
sneutrinos. The sneutrinos decay dominantly into neutrino and neutralino,
leading to a $\gamma \mett$ signal. We use identical cuts to that used for the
$\mu \gamma \mett$ channel, except for the cuts involving muons. 
Fig.~\ref{runIexpected}b
shows that between 5 and 35 events of this nature are expected within the
$95\%$ C.L. region. Standard Model backgrounds should be small, with dominant
physics background coming from $\gamma Z$ production, where $Z \rightarrow \nu
\bar\nu$. We emphasise that measuring this interesting channel would provide an
independent check on our model. 
In an R-parity conserving channel, a bound on the gravitino mass of $m_{\tilde
G}> 2.7 \times 10^{-5}$ eV~\cite{mangano} comes 
from the non-observation of signal $\gamma \mett$ in D0
data~\cite{dzero}. They place a bound of roughly 10 signal events for minimum
$E_T$'s of 25 GeV at the 95$\%$ C.L. for a luminosity of 13 pb$^{-1}$. This
would correspond to an upper bound of around 66 if we scale up to 86
pb$^{-1}$, as used here. The D0 bound is therefore not
restrictive\footnote{Also note that our cuts are completely different to those
in  the D0
analysis.}. The predicted rate anyway depends heavily upon the values of 
the unfitted parameters (see conclusions).

\FIGURE{%
\fourgraphs{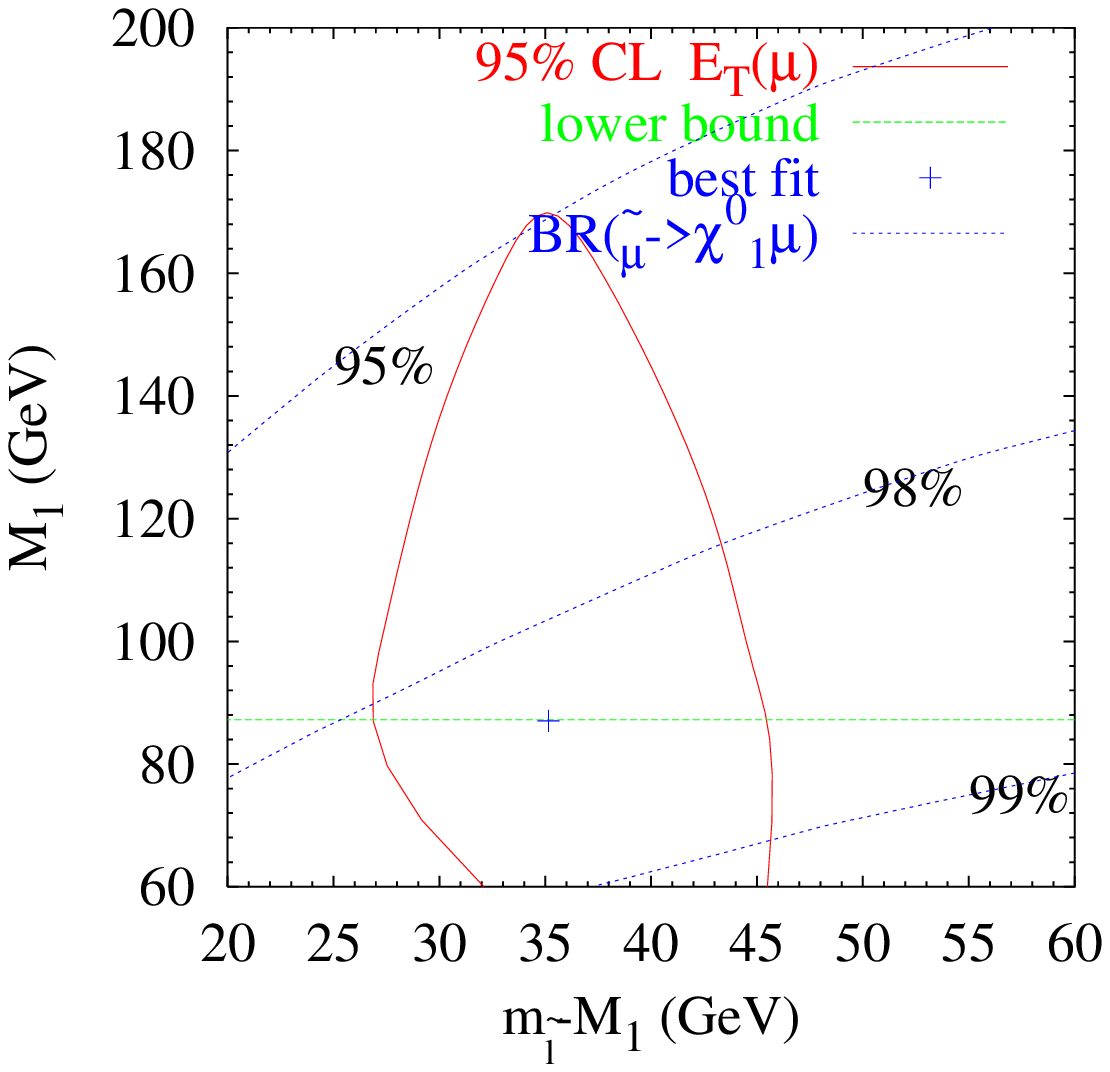}{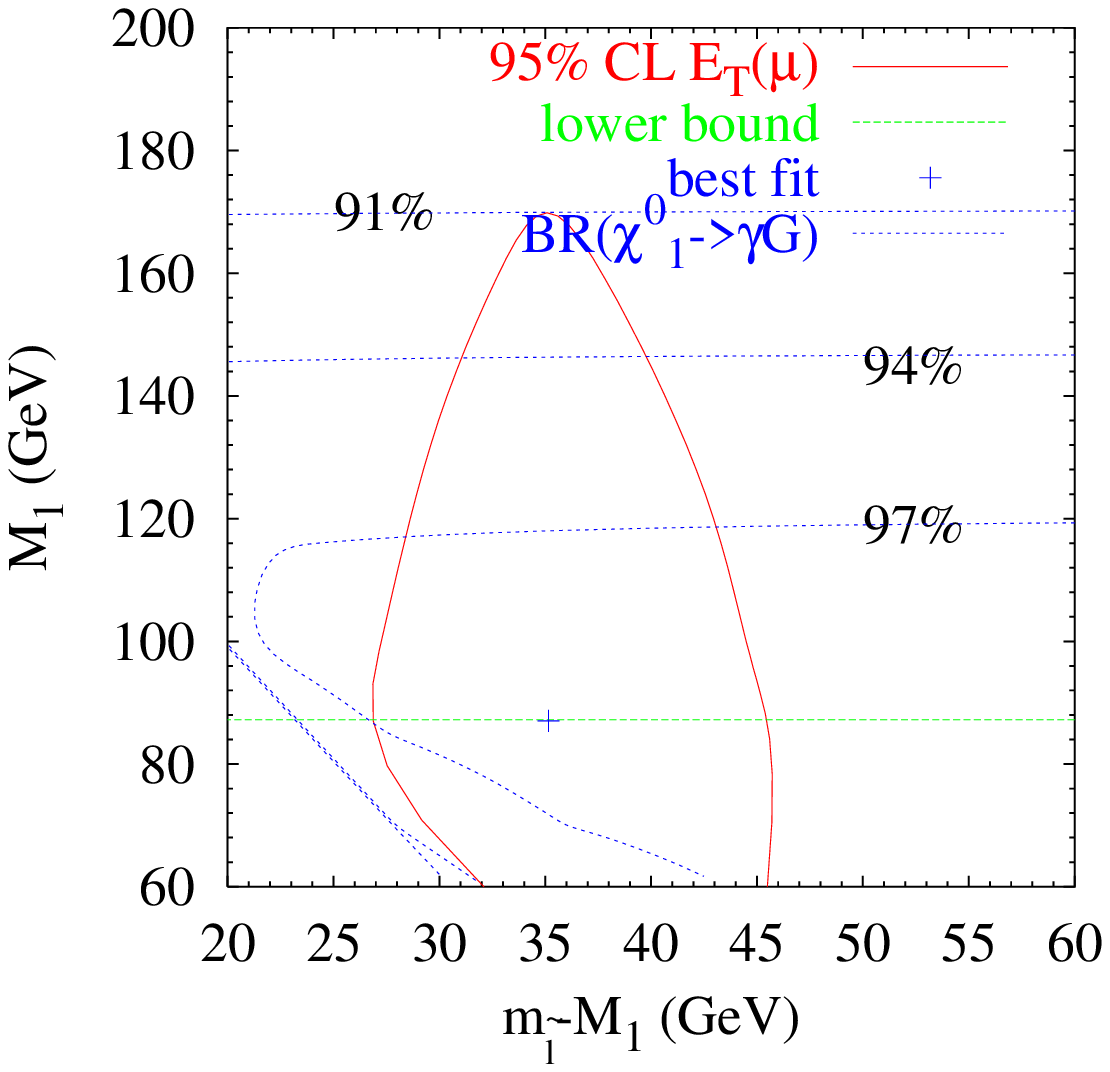}{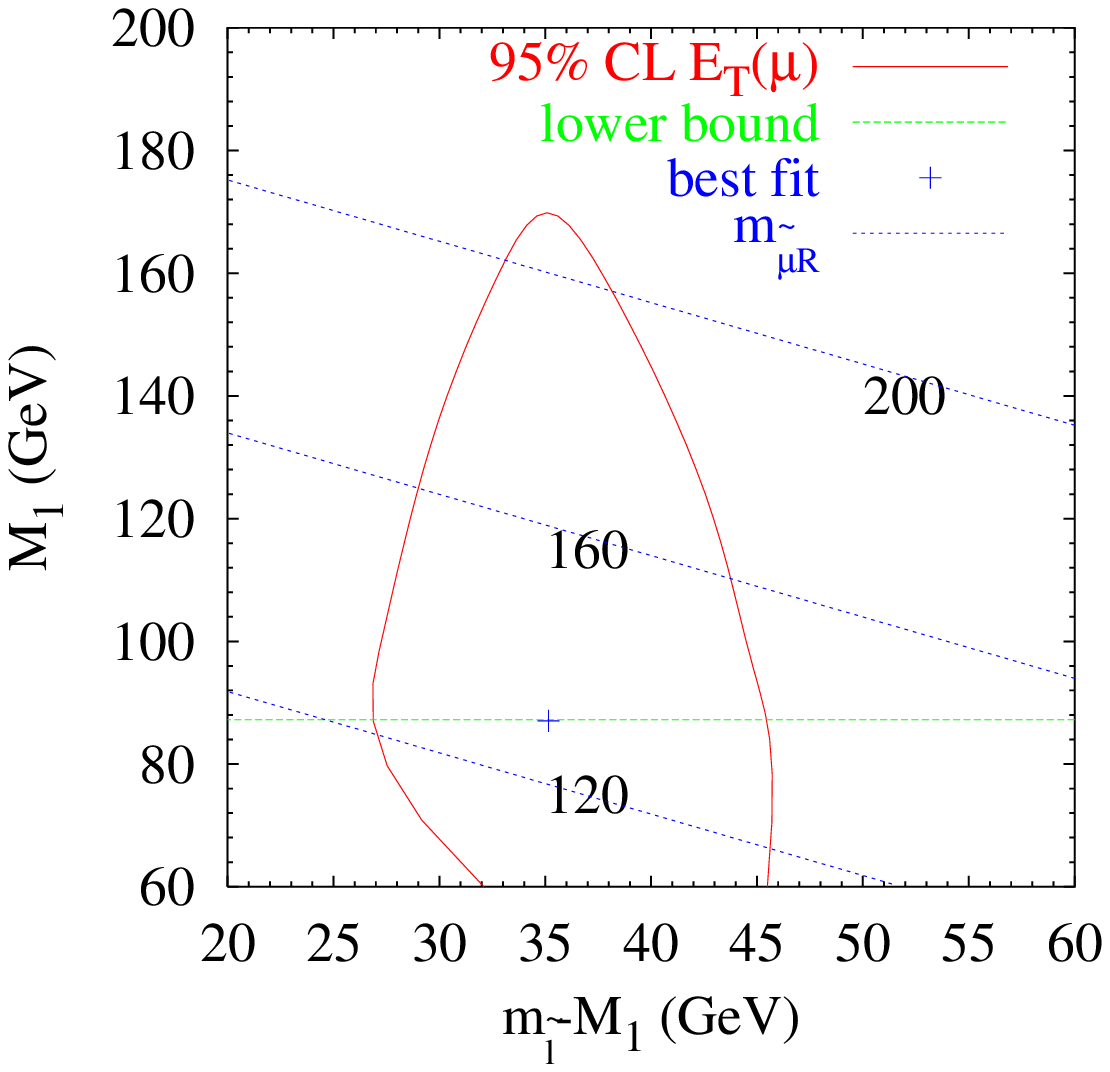}{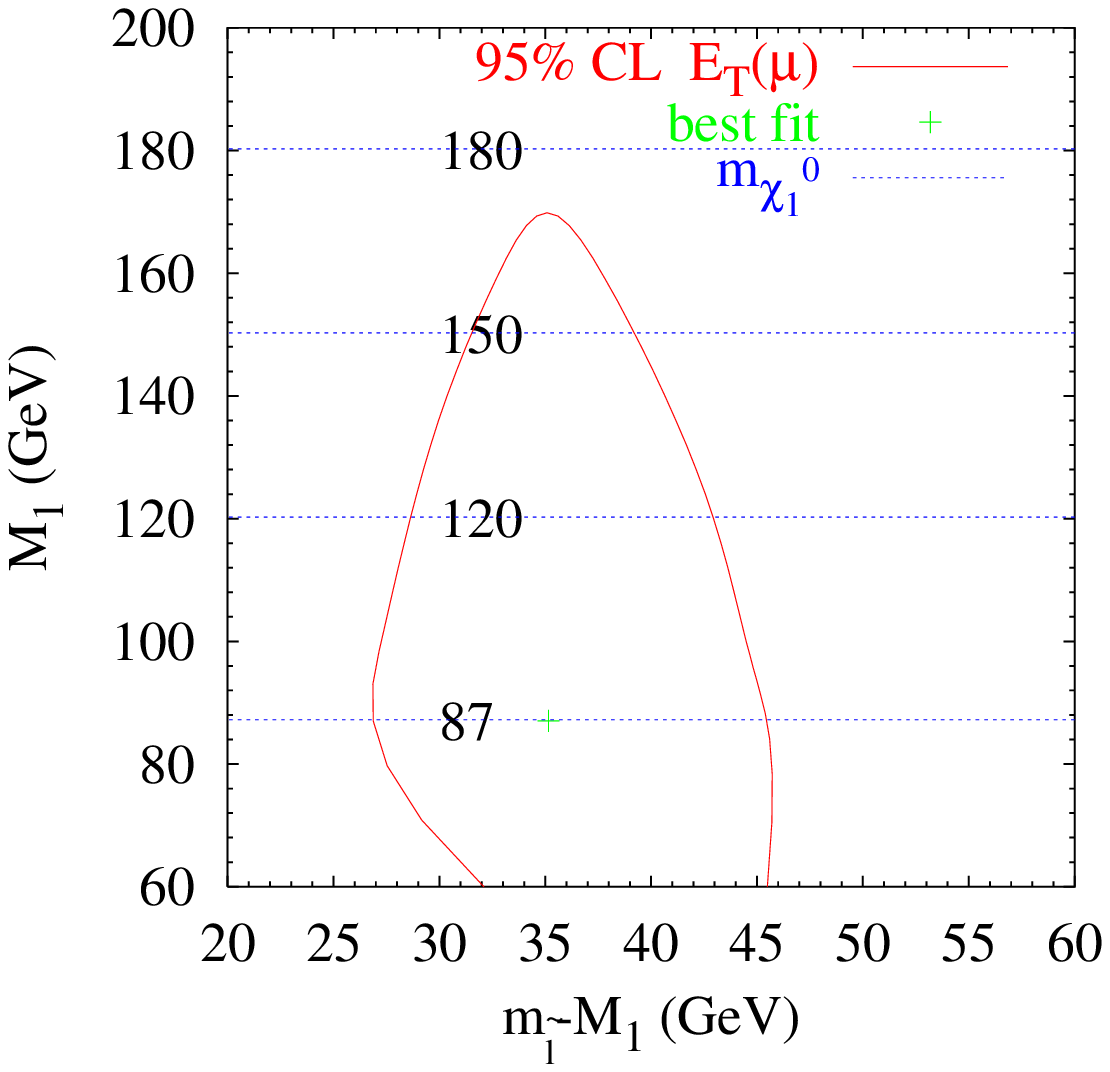}
\label{massBRs}
\caption{Relevant masses and branching ratios in the best-fit region.
The dashed blue lines show (labeled) contours of
(a) $BR({\tilde \mu} \rightarrow \mu \chi_1^0)$,
(b) $BR({\chi_1^0} \rightarrow {\tilde G} \gamma)$,
(c) $m_{{\tilde \mu}_R}$ (GeV),
(d) $m_{\chi_1^0}$ (GeV), the dotted green
line shows the lower bound coming from LEP2 and the red curve shows the 
region of good fit $E_T(\mu)$ in the for $\mu\gamma\mett$ events.}
}
In Figure~\ref{massBRs} we show the range of relevant masses and branching
ratios over the best-fit region. As Figure~\ref{massBRs}a indicates,
$0.95\leq BR({\tilde \mu} \rightarrow \mu \chi_1^0)<0.99$ in the best-fit region,
thus other decay modes of resonant smuon production ought to be suppressed. 
Similarly, $0.91\leq BR({\chi_1^0} \rightarrow {\tilde G} \gamma)<0.98$
thus the competing $\rpv$ (lepton and 2 jets) and $e^+ e^- {\tilde G}$ decay
modes of the resulting neutralino are also
suppressed to unobservable levels at Run I.
These branching ratios are dependent upon $\lambda'_{211}$ and $m_{\tilde G}$.
The range of viable right-handed smuon mass is $130 < m_{{{\tilde \mu}}_R} <
210$ GeV, as shown in Figure~\ref{massBRs}c. The lightest neutralino mass 
is approximately equal to $M_1$ and varies up to 170 GeV in the best-fit
region, as shown 
in Figure~\ref{massBRs}d.

\FIGURE{%
\threegraphs{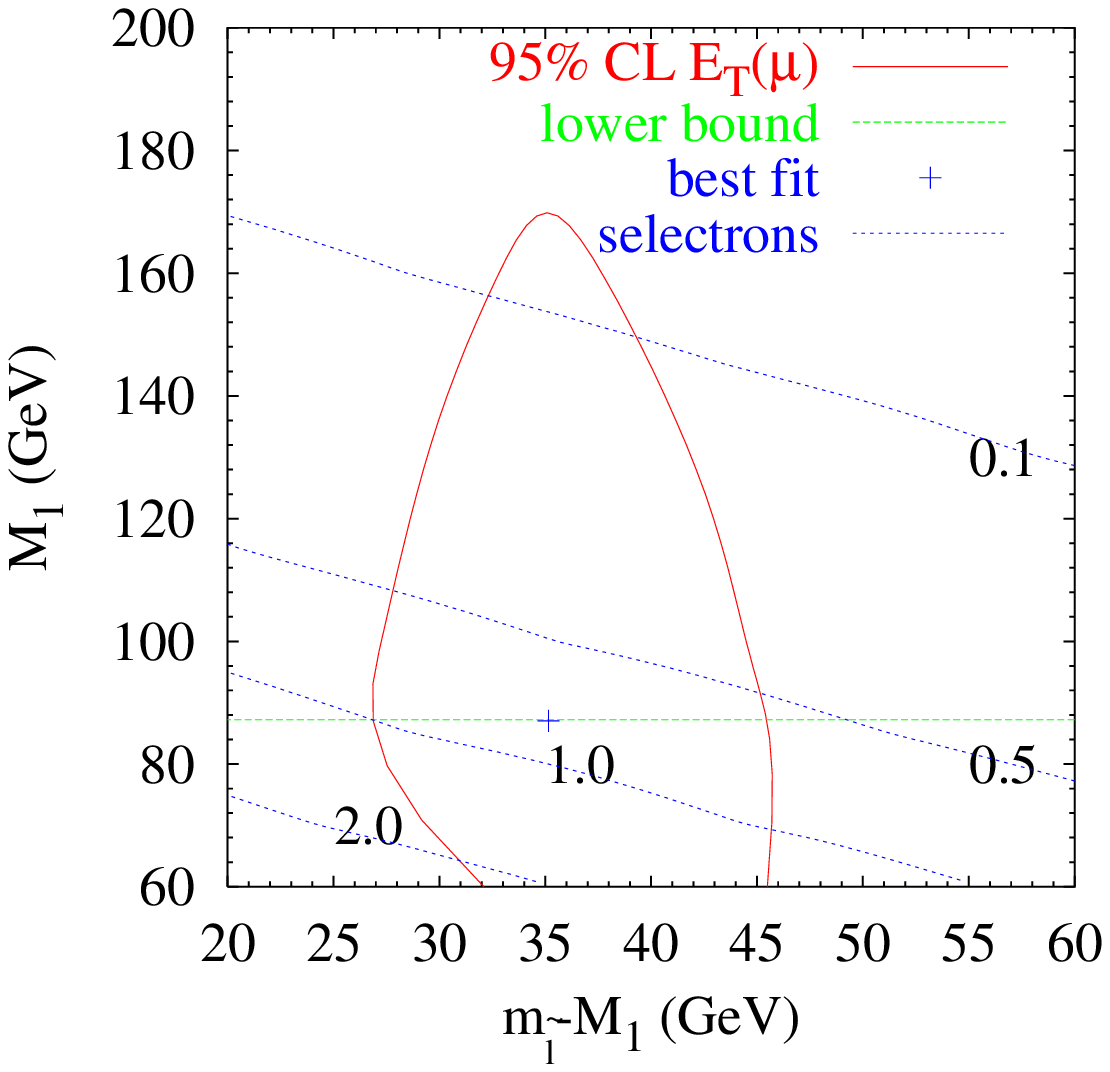}{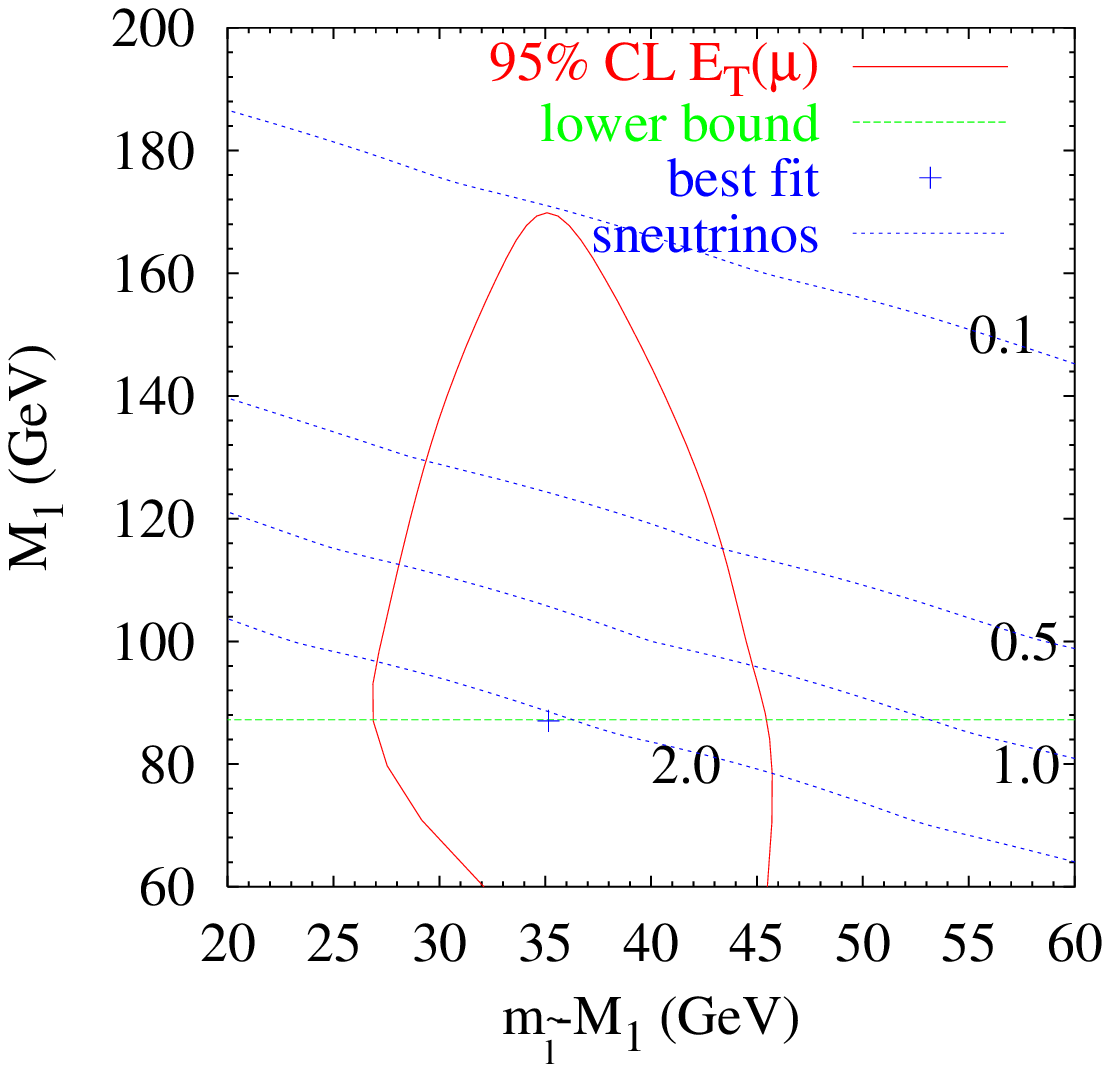}{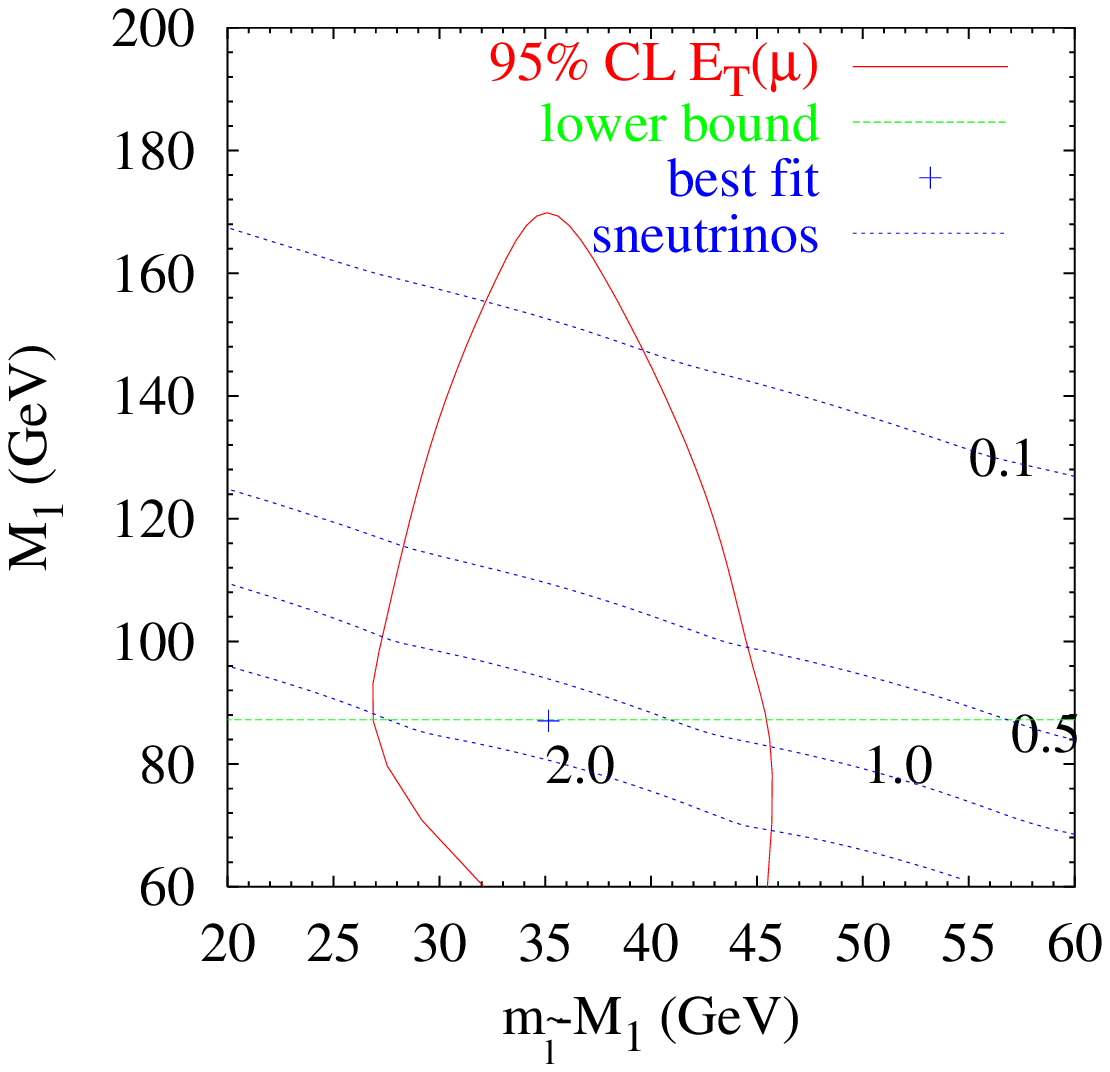}
\label{eeggmet1}
\caption{Number of (a) selectron, (b) selectron-sneutrino and (c) sneutrino
pairs produced at run I of the Tevatron. 
The dashed blue lines show the expected number of events, the dotted green
line shows the lower bound coming from LEP2 and the red curve shows the 
region of good fit $E_T(\mu)$ in the for $\mu\gamma\mett$ events.
}
}
The approximate level of other processes can be roughly estimated by
calculating 
the expected numbers of pairs of sparticles at Run I. Neutralino production is
predicted to be at an unobservable level, but the light sleptons have a
non-negligible number of expected pairs produced at Run I. Using
86.34 pb of luminosity, we display the expected number of selectron,
selectron-sneutrino and sneutrino pairs produced at Run I in
fig.~\ref{eeggmet1}. No experimental cuts at all have been applied to these
events, so detected numbers of these events might be expected to be a factor
of 5-10 less than the numbers that are displayed in the figure. We can see
from Figure~\ref{eeggmet1}a
that in the 95$\%$ C.L. region that fits the $E_T(\mu)$ distribution, there
are between about 0.1 and 1 selectron pairs expected, depending upon the
actual value of the parameters $\Delta m$ and $M_1$. 
The dominant decays of the selectrons is ${\tilde e} \rightarrow e
\chi_1^0$, again followed by ${\tilde \chi_1^0} \rightarrow \gamma {\tilde
G}$. Thus the selectron pairs provide the correct signature to describe 
the $ee \gamma \gamma \mett$ event recorded by CDF at run I. 
When detector
efficiencies taken into account, the expected number of $ee \gamma \gamma
\mett$ events, while being less than one, are nevertheless still much higher
than the expected Standard Model background. Selectron pair production is
predicted to be at the same level as smuon pair production, since they have 
approximately equal masses and they are produced via gauge interactions.
However, the muon detection efficiency is somewhat lower than for electrons,
so the expected number of {\em detected} $\mu \mu \gamma \gamma
\mett$ events from smuon pair production is smaller.
It is also possible to produce sneutrinos, and selectron-sneutrino pairs 
(with final state $e \gamma \gamma \mett$) are predicted to be at the 
0.1-2 event level before cuts.
In Figure~\ref{eeggmet1}c, we see that sneutrino pair production
is predicted to be at the level of 0.1 to 2 events. This final state will
manifest itself as $\mett \gamma \gamma$. 

\section{Predictions for Run II}
\FIGURE{%
\label{blahblah}
\twographs{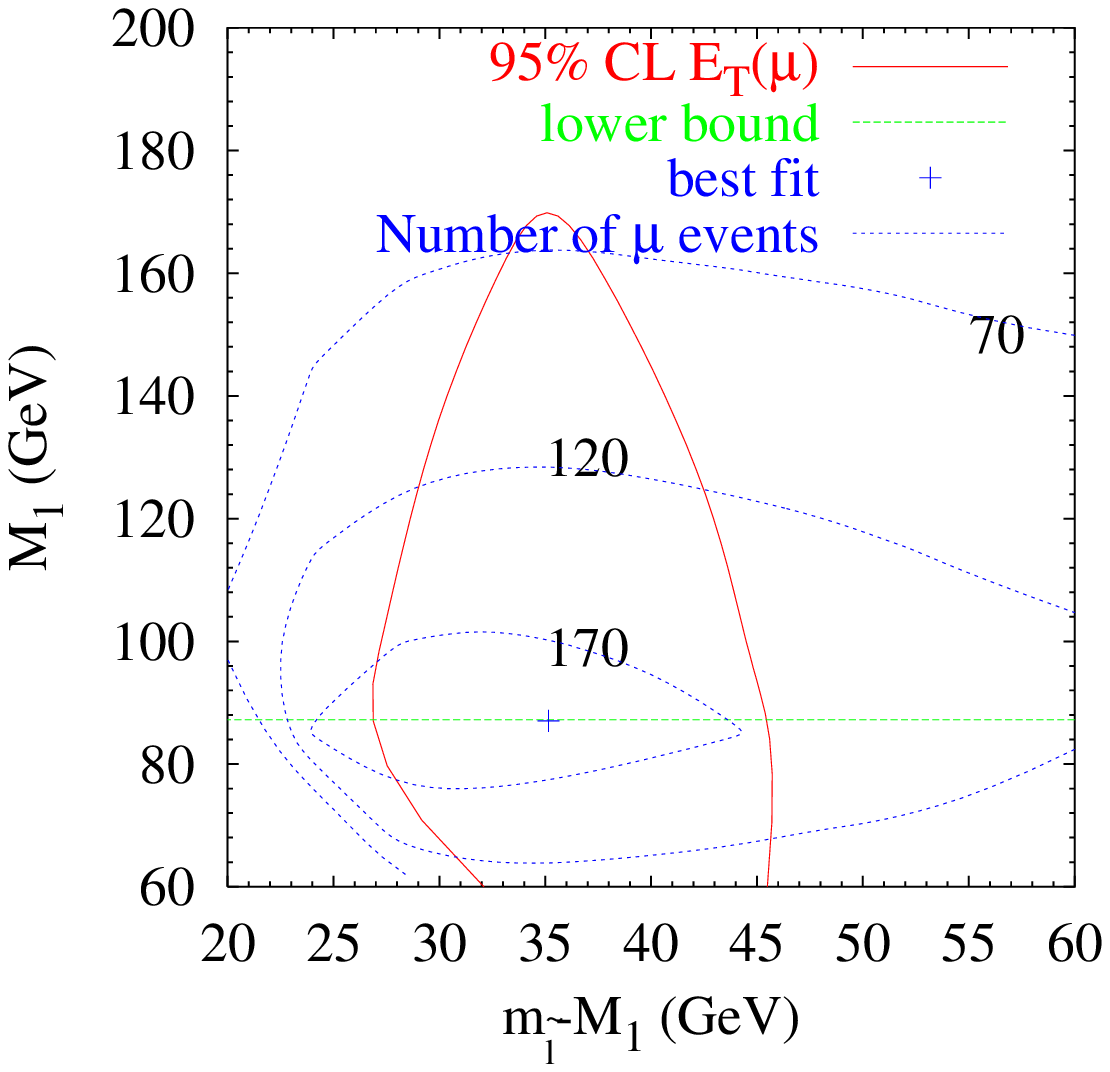}{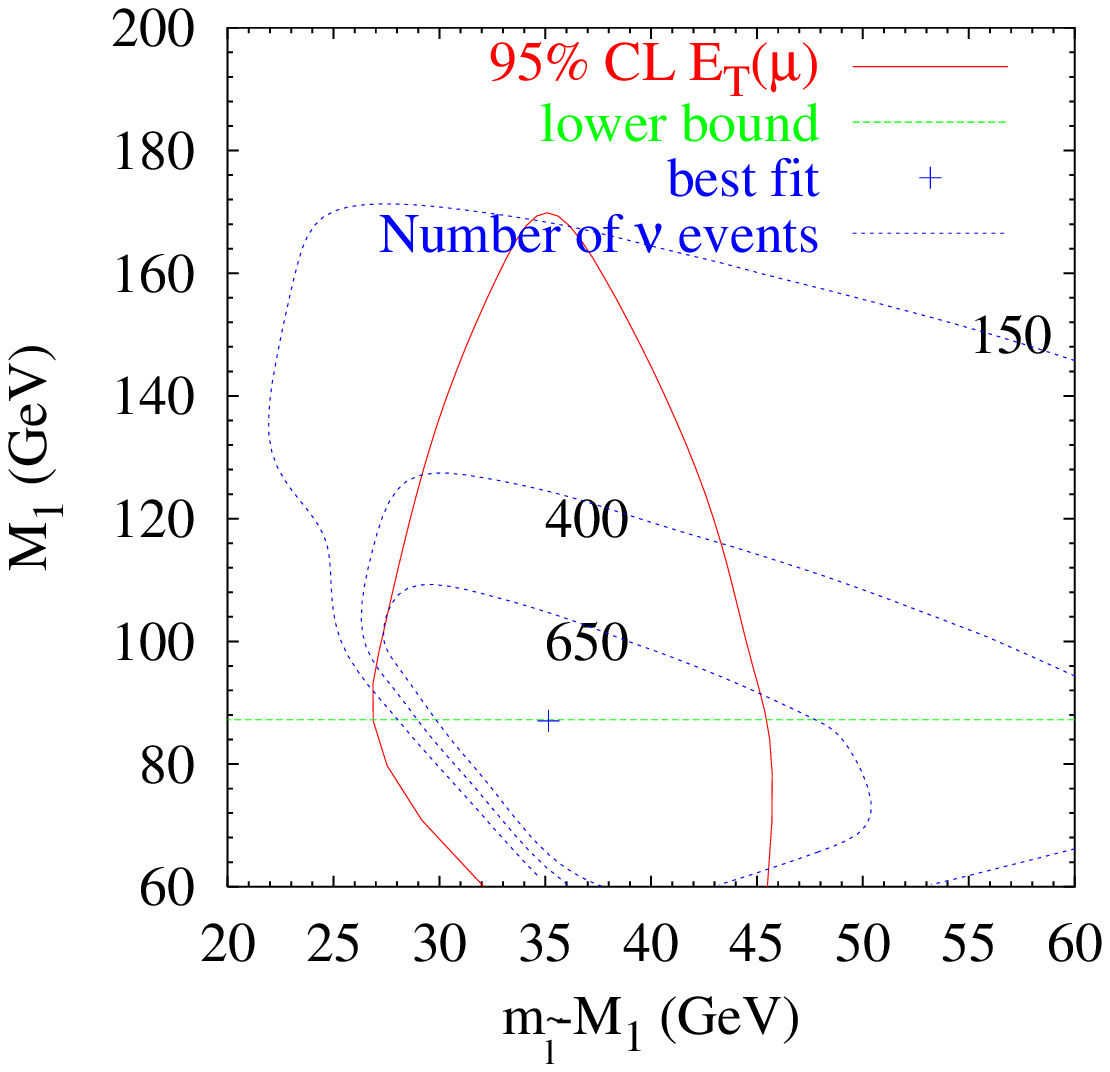}
\caption{Expected number of (a) $\mu \gamma \mett$ and (b) $\gamma \mett$
signal events in Run II data.
The dashed blue curves show (labeled) contours of number of expected events,
the dotted green
line shows the lower bound coming from LEP2 and the red curve shows the 
region of good fit $E_T(\mu)$ in the for $\mu\gamma\mett$ events.}
}
At Run II of the Tevatron, assuming 2 fb$^{-1}$ of luminosity, our model can
be ruled out or verified by again looking for an excess in the 
$\mu \gamma \mett$
channel, with much higher statistics. For example, assuming the same cuts and
detector efficiencies as in our Run I analysis, we expect 193
signal events at Run II for our best-fit point because the cross-section
increases to 0.096 pb for detected events. This estimate will be subject to
change once the relevant cuts and detector efficiencies for Run II are known.
We also expect an excess of 740 events in the $\mett \gamma$ channel
from resonant sneutrino production at the best-fit point. 
We calculate the number of events in the $\mett \mu \gamma$ and $\mett \gamma$
channel and display them in Figure~\ref{blahblah}, assuming 2 fb$^{-1}$ of
collected luminosity (assuming the same efficiencies and cuts as used in Run
I). We see that at least 70 events in the $\mett \mu \gamma$ channel are
expected and at least 150 in the $\mett \gamma$ channel. This will be
sufficient to measure parameters much more accurately, or rule the model out
completely. 

\FIGURE{%
\label{eeggmet4}
\threegraphs{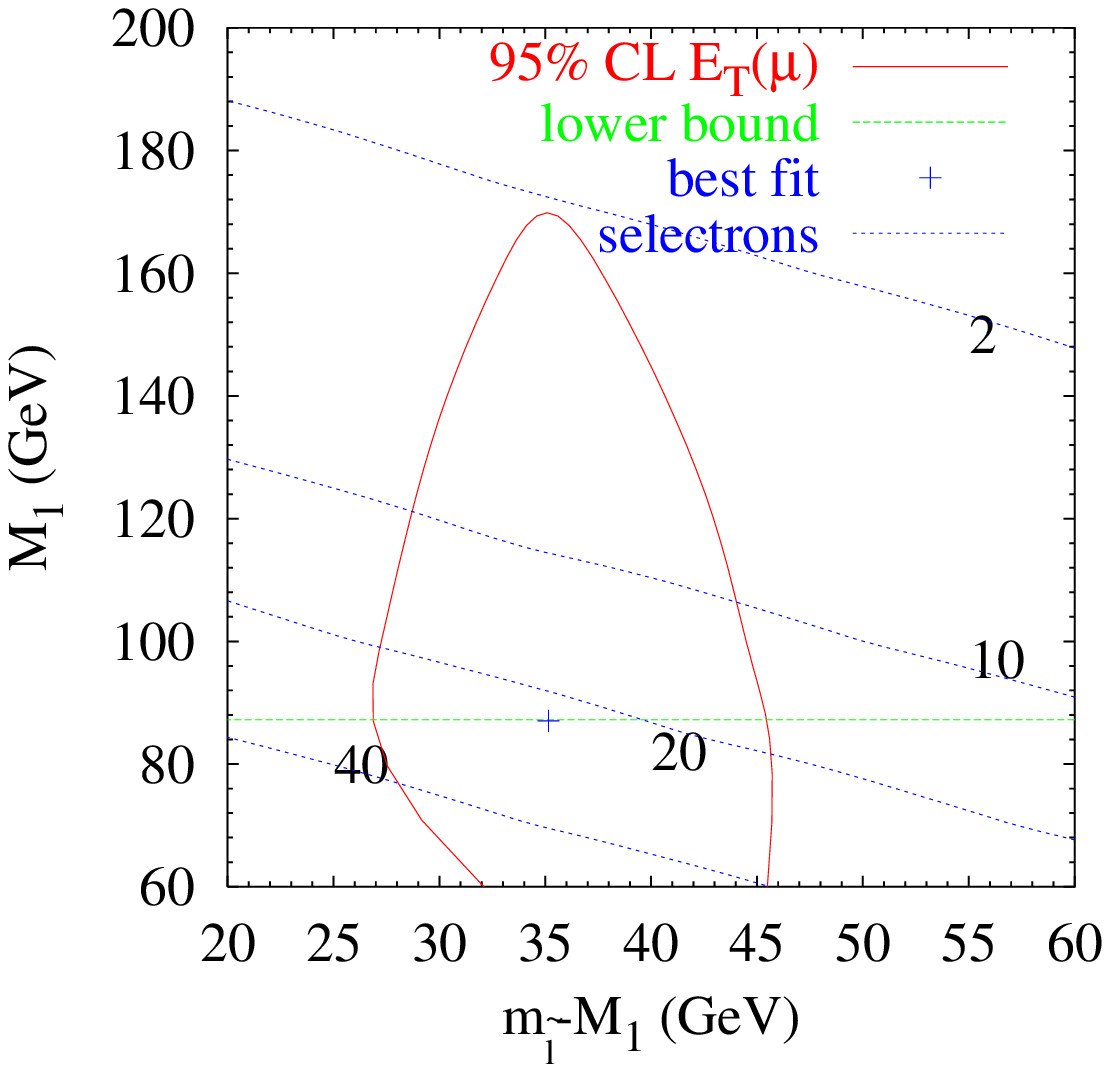}{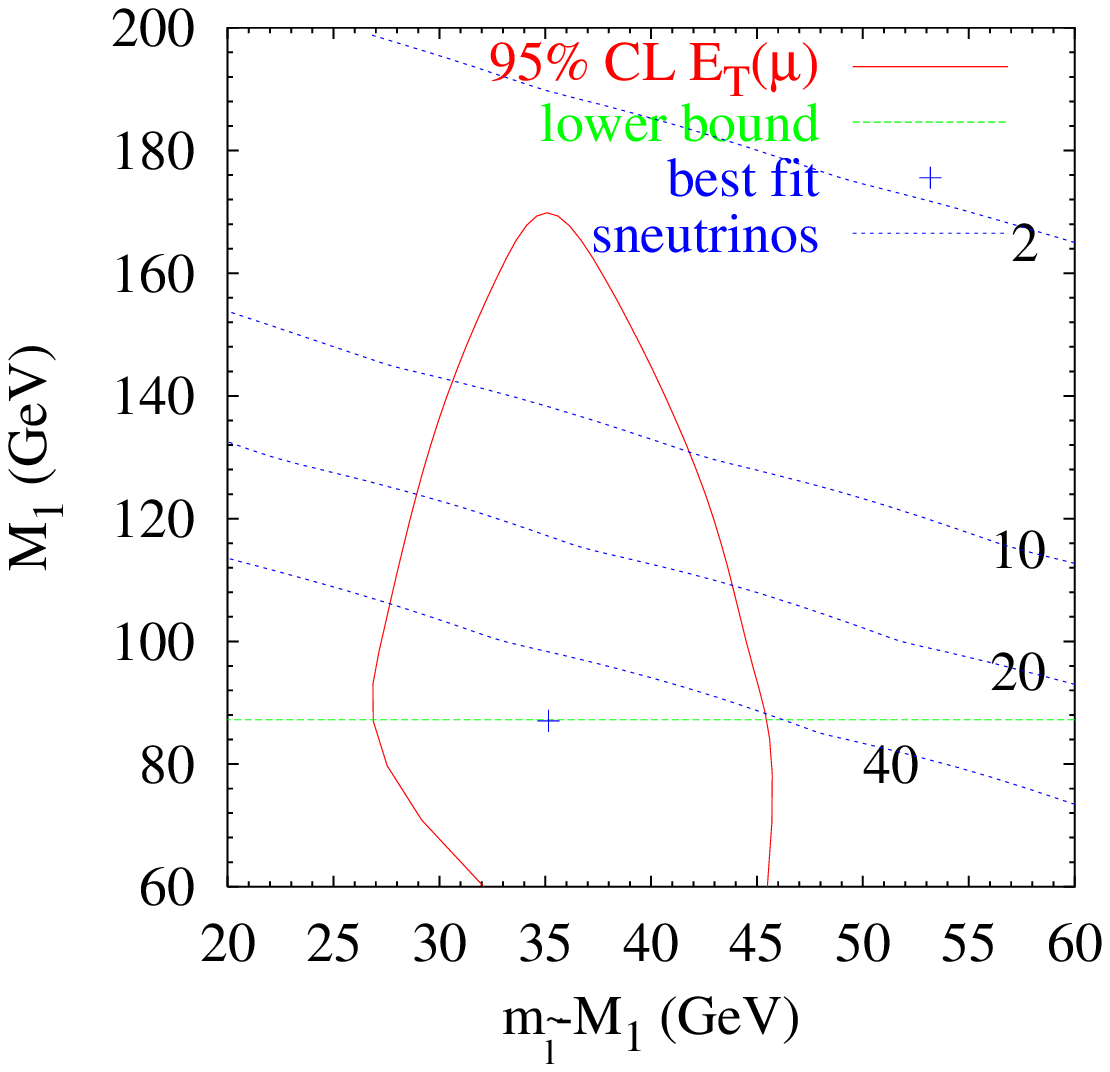}{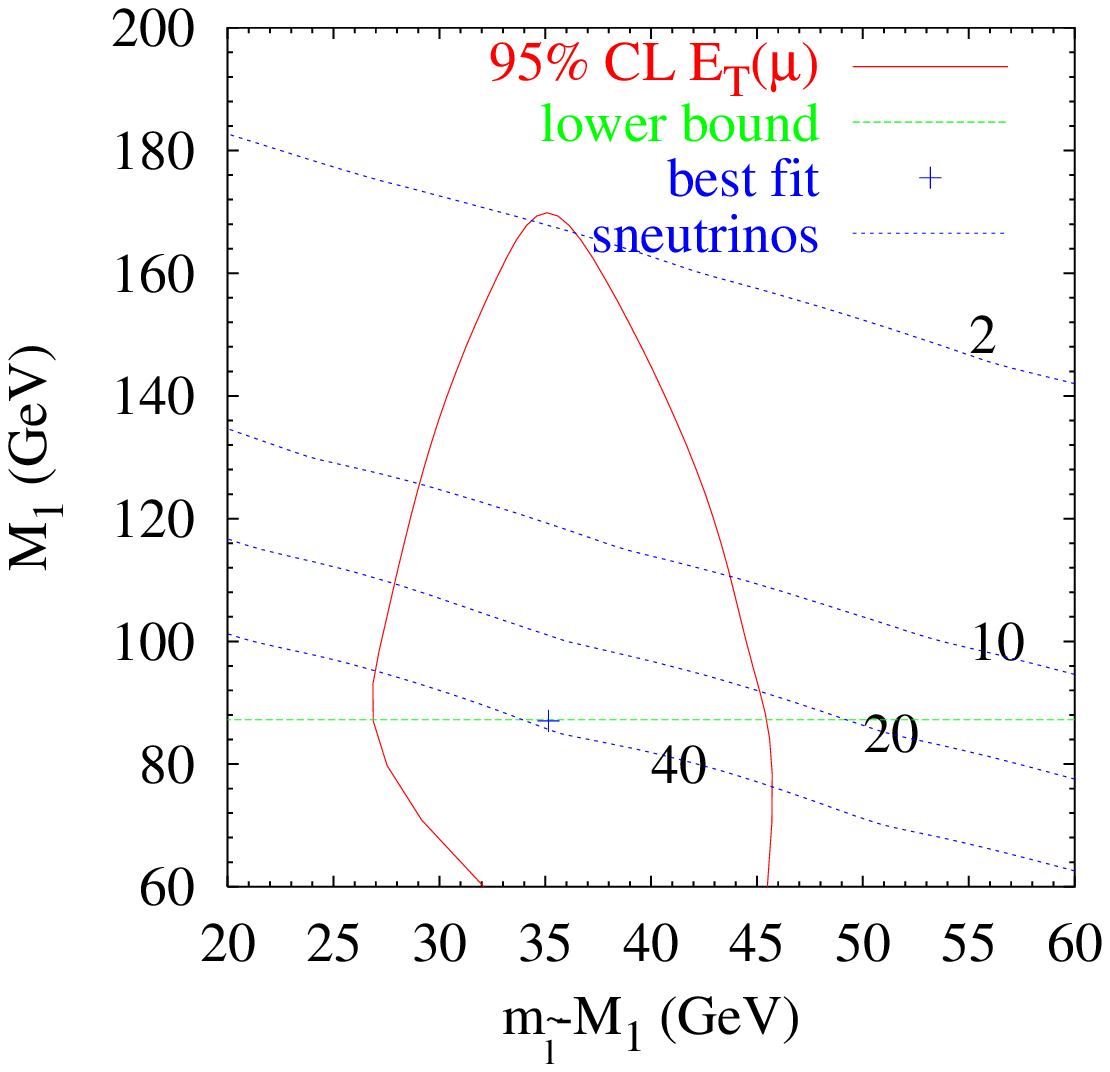}
\caption{Expected number of sparticle pairs produced at run II of the Tevatron
for 2 fb$^{-1}$ luminosity. 
The dashed blue lines show the expected number of events for (a) selectron
(b) selectron-sneutrino and (c) sneutrino pair production, the dotted green
line shows the lower bound coming from LEP2 and the red curve shows the 
region of good fit for $\mu\gamma\mett$ events.
}
}
If a $\mu \gamma \mett$ signal is seen at
Run II, the kinematic distributions will determine the viable
parameter space more accurately than Run I data. Sparticle pair production may
also be viable, and can provide constraints upon the parameter space. For this
reason, we show the expected number of slepton pairs produced in
Figure~\ref{eeggmet4}. Figure~\ref{eeggmet4}a shows that between 2 and 30
selectron (and smuon) pairs are expected, between 3 and 50 sneutrino-selectron
and sneutrino-smuon pairs (Figure~\ref{eeggmet4}b), and between 2 and 40
sneutrino pairs (Figure~\ref{eeggmet4}c). Once efficiencies and cuts are taken
into account, this adds up to at most a handful of events in each
channel. Nevertheless, this would provide independent confirmation for our
scenario and Standard Model background rates would be still extremely low.

\section{Conclusions}

We have demonstrated that R-parity violating supersymmetry with a 
light gravitino can explain an anomalously high measured cross-section for
the $\mu \gamma \mett$ channel. 
It also explains features observed in the kinematic variables of the signal
events. We have used this information to constrain the slepton and neutralino
mass parameters in the model. 
Whereas we could not perform a combined fit to all the different kinematic
distributions, if $M_1 \approx 90$ and $\Delta m_{\tilde l}\approx 30$ GeV,
all 
of the distributions are fit well. Ideally a combined fit to all kinematical
variables would be performed for our model and a measure of the fit
probability calculated 
(along with that of the Standard Model). Such a fit would require
simulation of the backgrounds as well as knowledge of the multi-dimensional
distributions of variables rather than the one-dimensional projections
available to us.
We have seen qualitatively that the $ee \gamma \gamma \mett$ event observed in
Run I is fit much better by our model than by the Standard Model.

We chose representative parameters $m_{\tilde G}=10^{-3}$ eV, 
$\lambda'_{211}=0.01$ and $\tan \beta=10$. Varying the first two parameters
does not change the kinematics of the event, merely the branching ratios of
the decays. Thus the total number of signal events in the relevant channel
changes, but the kinematic shapes in the signal events remain the same. 
A higher value of $m_{\tilde G}$ decreases the number of neutralinos decaying
to a photon and a gravitino, but this can be compensated for by increasing
$\lambda'_{211}$ to increase the production cross-section. However, at some
value of $\lambda'_{211}$, R-parity violating decays will dominate over the
gravitino decays of the neutralino. We also note that changing 
$\tan \beta$ has the effect of changing the relationship between
$m_{\tilde l}$ and the slepton masses, as eq.~\ref{slepmasses} shows. Thus,
different 
values of $\tan \beta$ could potentially prefer different regions of $\Delta
m$. 
\TABULAR{c|cccc}
{\label{tab:pars}
Quantity  & 1 &2  &3  &4  \\ \hline
$\lambda'_{211}$ & 0.02 &0.01 & 0.01& 0.03\\
$m_{\tilde G}$ (eV)& 0.01& $10^{-3}$&$10^{-3}$ &$10^{-3}$ \\
$\tan \beta$ &10  &20 &5 &10 \\
$\Delta m$&35 &40 & 35& 35\\
$M_1$   & 87& 87& 87& 200  \\ \hline
$N_{\mett \mu \gamma}$& 8.0& 6.7& 7.9& 10.7 \\
$N_{\mett \gamma}$ & 42& 34& 34& 14\\ \hline
$\Delta\sigma(E_T(\mu))$                   & 3.3& 3.1& 3.3& 3.3  \\
$\Delta\sigma(E_T(\gamma))$   		   & 1.8& 1.9& 1.9& -1.1 \\
$\Delta\sigma(\mett)$                      & 2.2& 2.4& 2.4& 1.0  \\
$\Delta\sigma(m_{\mu \gamma})$             & 2.6& 2.3& 2.6& -1.5 \\
$\Delta\sigma(M_T(\mett\mu))$              & 2.1& 2.5& 2.3& 1.9  \\
$\Delta\sigma(M_T(\mett\gamma))$           & 2.8& 3.0& 2.9& 0.6  \\
$\Delta\sigma(M_T( \mett\gamma \mu))$      & 1.7& 1.6& 2.0& 1.0  \\
$\Delta\sigma(\Delta \phi_{\mett \gamma })$& 2.0& 2.2& 2.1& 2.1  \\
$\Delta\sigma(\Delta \phi_{\mu \gamma})$   & 2.7& 2.7& 2.8& 2.6  \\ 
$\Delta\sigma(\Delta \phi_{\mett \mu})$    & 2.8& 2.7& 2.8& 3.1  \\
$\Delta\sigma(H_T)$                        & 1.8& 2.1& 2.1& 1.4  \\
$\Delta\sigma(\Delta R_{\mu \gamma})$      & 2.3& 2.3& 2.4& 2.7  \\
}
{Examples of 4 different parameter points, showing the parameters, the number
of predicted events ($N_{\mett \mu \gamma}, N_{\mett \gamma}$) and the number
of $\sigma$ that the point fits the data {\em better} than the Standard Model.}
Higher values of $\lambda'_{211}$ can also be accommodated by increasing both
the mass of the sleptons and the mass of the neutralino to decrease the hard
production cross-section. 
These points are illustrated in turn in table~\ref{tab:pars}, where the parameters
are 
all varied to have in such a way as to produce a number of events comparable
to the best fit value of 7.8. Point 1 illustrates that a higher value for
$m_{\tilde G}$, which gives lower branching ratios of $\chi_1^0 \rightarrow
\gamma {\tilde G}$, can be compensated by an increase in $\lambda'_{211}$,
raising the resonant smuon production cross-section. 
The kinematic fits still look favourable: point 1 fits the data better than
the Standard Model to $\Delta \sigma=1.7-3.3$ depending upon the variable.
However, it does seem
difficult to raise $m_{\tilde G}$ further because the required increase in 
$\lambda'_{211}$ then gives dominant decay modes to be R-parity violating,
decreasing the number of signal events $N_{\mett \mu \gamma}$.
Points 2 and 3 illustrate the insensitivity to
values of $\tan \beta$. We note in point 4 that a heavy neutralino of 200 GeV
also can provide enough events by raising $\lambda'_{211}$, but that the
$E_T(\gamma)$ distribution is predicted to be too hard compared with the
data. Also, the signal bump in $m_{\mu \gamma}$ goes to higher energies,
disagreeing somewhat with the data. The other distributions seem to fit the
data quite well however. 
Point 4 also shows that the predicted number of
$\gamma \mett$ events does vary with $\lambda'_{211}$ and $m_{\tilde G}$, and
so our 
prediction for the numbers of these events is parameter dependent.

Run II will provide a definitive test of our scenario,
by again looking in the $\mu \gamma \mett$ and $\mett \gamma$ channels. 
Other final states with small SM backgrounds are expected at the
few-event level. For example, observation of $ee \gamma \gamma \mett$, $\mu
\mu \gamma \gamma \mett$, $\mu \gamma \gamma \mett$ and $e \gamma \gamma \mett$
would provide independent confirmation of our scenario. We note, however,
that towards the higher values of $M_1$ expected, less than one event in each
of the pair production channels is likely once detector effects and cuts are
taken into account.

\acknowledgments
We would like to warmly thank H. Frisch and D. Toback for their gracious
help and advice regarding the data. We would like to thank
D.Hutchcroft for discussions concerning LEP data and A. Barr, G. Blair and
J. Holt for discussions on the statistics. We also acknowledge discussions
with G. Altarelli, J. Ellis and M. Mangano.

%======================================================================

\end{document}